\documentclass[a4paper,11pt]{article}
\pdfoutput=1 %
\usepackage{soul}

\usepackage{jcappub} %

\usepackage[T1]{fontenc} %
\usepackage[normalem]{ulem}
\usepackage{float}

\title{GEO-FPT: a model of  the galaxy bispectrum at mildly non-linear scales}
\author[a,b]{Sergi Novell-Masot,}
\author[a]{Davide Gualdi,}
\author[a,b]{H\'ector Gil-Mar\'in,}
\author[a,d]{Licia Verde}%

\affiliation[a]{ICCUB, University of Barcelona, Mart\'i i Franqu\`es, 1, E-08028 Barcelona, Spain}
\affiliation[b]{Institut d’Estudis Espacials de Catalunya (IEEC), Barcelona E08034, Spain}
\affiliation[d]{ICREA, Pg. Llu\'is Companys 23, Barcelona, E-08010, Spain}

\emailAdd{sergi.novell@icc.ub.edu}

\abstract{We present GEO-FPT (Geometric Fitted Perturbation Theory), a new model   for the galaxy bispectrum anisotropic signal in redshift space, with functional form rooted in perturbation theory. It also models the dependence of the bispectrum with the geometric properties of the triangles %
 in Fourier space, and has a broader regime of validity than  
state-of-the-art theoretical models based on perturbation theory. We calibrate the free parameters of this model using high-resolution dark matter simulations and perform stringent tests to show that  GEO-FPT 
describes the galaxy bispectrum accurately up to scales of $k\simeq0.12 h{\rm Mpc}^{-1}$ for different cosmological models, as well as for biased tracers of the dark matter field, considering a survey volume of $100$ (Gpc $h^{-1})^3$. In particular,  a joint analysis of 
the power spectrum and bispectrum anisotropic signals, taking into account their full covariance matrix, reveals that  the 
relevant physical quantities -- 
the BAO peak position (along and across the line-of-sight), and the growth of structure parameters times the amplitude of dark matter fluctuations, $f\sigma_8$-- are recovered  in an unbiased way, with an accuracy better than  $0.4\%$ and  $2\%$ respectively (which is our $2\sigma$ statistical limit of the systematic error estimate). 
In addition, the bispectrum  signal breaks the $f\sigma_8$ degeneracy without detectable bias: $f$ and $\sigma_8$ are recovered with  better than 2.7\% and 3.8\% accuracy respectively (which is our $2\sigma$ statistical limit of the systematic error estimate).

GEO-FPT boosts the applicability of the bispectrum signal of galaxy surveys beyond the current limitation of $k\lesssim 0.08\,h$  Mpc$^{-1}$ %
and makes the bispectrum a  key  statistic to unlock the information content from the mildly non-linear  regime in the on-going and forthcoming galaxy redshift surveys.}

\begin{document}

\maketitle

\section{Introduction}
\label{sec: intro}
The power spectrum has long been the primary statistic used for constraining the cosmological parameters and the evolution of perturbations. This is a crucial component for the development of the current cosmological standard model, the $\Lambda$CDM model. However, the power spectrum alone would only be sufficient in the case of the description of a purely Gaussian field where, according to Wick's theorem, all the information of the field is enclosed in the two-point statistics. This is not the case for the galaxy over-density field of the evolved universe, with gravitational collapse (and possibly even primordial fluctuations) being sources of non-Gaussianity. 

Hence, higher-order correlations, with the bispectrum as the most straightforward one, can yield further information about the evolution of structure in the universe. Specifically, the bispectrum  of large-scale structure surveys can lift degeneracies between cosmological parameters, thus improving the precision in their inference \cite{Verde1998,Matarrese1997,Sefusatti:2006pa,fry1993biasing,Ruggeri_2018,Hahn_2020,oddo2021cosmological,bartolo2013matter,bellini2015signatures,bertacca2018relativistic,Gil_Mar_n_2011,yankelevich2019cosmological,coulton2019constraining,ruggeri2018demnuni}, especially if the quadrupoles are added \cite{gagrani2017information,Gualdi:2020ymf,rizzo2023halo,ivanov2023cosmology,d2022one}. Moreover,  it is the lowest order  correlator which   encloses  primordial non-Gaussianity information  \cite{baldauf2011primordial,dizgah2021primordial, gualdi_matter_2021}.  

Although the bispectrum has been studied for decades, it has not been applied as widely as the power spectrum
to real survey data 
(see \cite{Scoccimarro01,2dFbispectrum,Gil-Marin:2016wya} as pioneering works on extracting information from the galaxy bispectrum).
This is both because of the inherent complexity of the bispectrum formalism, and because  of the %
stringent  requirements on control of systematics and on survey size that 
are needed %
to confidently measure the bispectrum, and  which have only been %
achieved  relatively recently (BOSS  and eBOSS \cite{BOSS, Gil-Marin:2016wya, eBOSS,eBOSS2}).

Current and forthcoming surveys (such as DESI\footnote{\url{http://desi.lbl.gov}} \citep{Levi:2013gra}, Euclid\footnote{\url{http://sci.esa.int/euclid/}} \citep{Laureijs:2011gra}, Vera Rubin\footnote{\url{https://www.lsst.org/}} \citep{Abell:2009aa}, Roman\footnote{ \url{https://roman.gsfc.nasa.gov}} \cite{Green:2012mj}) will reach an unprecedented degree of systematics suppression and large survey volumes, making these surveys particularly well-suited to study the bispectrum and other higher-order correlations. It is thus essential to accurately model these summary statistics, in order to obtain reliable results as precisely as possible.

This paper presents a phenomenologically motivated effective bispectrum model in redshift space, the `GEO-FPT' model, that describes the bispectrum monopole and quadrupoles. It consists in a phenomenological  modification of the  the perturbation theory $Z_2$ kernel with 5 free parameters, which can vary across redshift. The free parameters are calibrated  with bispectrum measurements of the \textsc{Quijote} suite of simulations \cite{villaescusa-navarro_quijote_2020}. This results in a fit which achieves a $\leq 3\%$ accuracy for most of the configurations of the bispectrum monopole, and of $\lesssim 30\%$ for the quadrupoles, for triangles whose sides fulfil $0.02\leq k\leq0.12$ $h\,\textrm{Mpc}^{-1}$ at  $z=0.5,\,1,\,\textrm{and}\, 2$.

This approach 
differs from other techniques such as  \cite{angulo2015one,ivanov2022precision}, where the modifications from SPT come in the form of counterterms that are marginalized over in the cosmological parameter inference. 
In GEO-FPT, which is similar to the models proposed by \cite{Scoccimarro2001, GilMarin:2011ik, Gil-Marin:2014pva}, once the best-fit  values for the free parameters are obtained, they remain fixed, assuming that the cosmology dependence of the $Z_2$ kernel and the form of the kernel's corrections is weak. This assumption is  validated on  simulations with different cosmologies.

GEO-FPT's regime of validity reaches  mildly non-linear scales,
achieving a significant precision gain with respect to  power spectrum-only analyses. We also present a battery of tests which attest that GEO-FPT  can be successfully used on the upcoming surveys to exploit the information contained in the bispectrum statistic. 

This paper is structured as follows: In Section \ref{sec: theory} we provide a brief overview of  the perturbation theory models for power spectrum and bispectrum, as well as our new effective model. In Section \ref{sec: methodology} we present the simulations and the methodology, and in Section \ref{sec: results} we show our main results for the GEO-PT model. In Section \ref{sec: conclusions} we summarize our main findings, and discuss the implications of our work for future surveys and experiments.  

\section{Theoretical modelling}
\label{sec: theory}
We start by reviewing the standard perturbation theory formalism for power spectrum and bispectrum in real and redshift space. The shortcomings of this description will motivate the theoretical model we propose. 

The real space power spectrum $P_{\delta\delta}$ (which we will use as input for computing the bispectrum) is obtained at 2-loop using the \textsc{PTcool} code\footnote{\url{https://github.com/hectorgil/PTcool}}. In the standard perturbation theory (SPT) framework, it can be expressed at two loops as
\begin{equation}   
\label{eq: Pnl_matter}P_{\delta\delta}=P^L+2P_{\delta\delta}^{(13)}+P_{\delta\delta}^{(22)}+2P_{\delta\delta}^{(15)}+2P_{\delta\delta}^{(24)}+2P_{\delta\delta}^{(33)},
\end{equation}
where $P^L$ is the linear power spectrum, obtained with \textsc{Class} \cite{Lesgourgues:2011re}, and the terms $P_{\delta\delta}^{(ij)}$ are loop correction terms, whose explicit expression can be found e.g. in  \cite{2002PhR...367....1B} (the dependence on $k$ is implicit). In \cite{Crocce_2006}, a modification of the SPT formalism was proposed, denoted renormalized perturbation theory (RPT), which, leaving formally Equation \ref{eq: Pnl_matter} unchanged, amends the expressions for $P_{\delta\delta}^{(ij)}$ and improves the accuracy of the SPT power spectrum in the mildly non-linear regime \cite{Gil_Mar_n_2012}. We will use this approach at two loops (2L-RPT) throughout this work. %

The simplest  model for bispectrum in real space, that uses perturbation theory at tree level, is referred to as the standard perturbation theory (SPT) model, and  can be written as \cite{Fry1983}:
\begin{equation}
\label{eq: bisp_spt}
B(\textbf{k}_1,\textbf{k}_2,\textbf{k}_3)=2F_2^{\textrm{SPT}}(\textbf{k}_1,\textbf{k}_2)P^L(\textbf{k}_1)P^L(\textbf{k}_2)+2\textrm{ perm.},
\end{equation}
where $F_2^\textrm{SPT}$  is the second order (SPT) kernel, given by
\begin{equation}
F_2^{\textrm{SPT}}(\textbf{k}_1,\textbf{k}_2)=\frac{5}{7}+\frac{1}{2}\cos(\theta_{12})\left(\frac{k_1}{k_2}+\frac{k_2}{k_1}\right)+\frac{2}{7}\cos^2(\theta_{12}),
\end{equation}
where we note as $\theta_{12}$ the angle between $\textbf{k}_1$ and $\textbf{k}_2$. 

The redshift space\footnote{Here and hereafter we assume flat sky, distant observer approximation.} galaxy power spectrum, $P_g$, is calculated from both the non-linear matter power spectrum, $P_{g,\delta\delta}$, the density-velocity, $P_{g,\delta\theta}$, and velocity-velocity, $P_{g,\theta\theta}$, power spectra, according to the TNS model \cite{Taruya_2010,Nishimichi_2011},
\begin{align}
\label{eq: Pnl_redshift}
     P_g(k,\mu)&=D_\textrm{FoG}^P(k,\mu,\sigma_P)\big[P_{g,\delta\delta}(k)+2f\mu^2P_{g,\delta\theta}(k)+f^2\mu^4P_{\theta\theta}\nonumber\\
     &+b_1^2A^\textrm{TNS}(k,\mu,f/b_1)+b_1^4B^\textrm{TNS}(k,\mu,f/b_1)\big],
\end{align}
where $f$ denotes the  logarithmic growth rate of perturbations, and $d \ln \delta /d\ln a$ and  $P_{g,\delta\delta}, P_{g,\delta\theta}$ are computed as in \cite{beutler2014clustering}. In doing so, we are using the bias expansion $\{b_1,b_2,b_{s^2},b_\textrm{3nl}\}$ and assuming the Lagrangian local bias approximation, which has been validated on simulations for haloes \cite{baldauf2012evidence,saito2014understanding}, as well as BOSS-type galaxies \cite{Brieden_ptchallenge}. In Equation \ref{eq: Pnl_redshift}, $A^\textrm{TNS},B^\textrm{TNS}$ are functions defined in \cite{Taruya_2010}, $\mu$ is the cosine of the angle of $k$ with the line of sight, and $D_\textrm{FoG}^P$ is a damping factor that accounts for the Fingers-of-God (FoG) effect of  redshift space distortions \cite{Jackson_1972}. For the power spectrum, we model it as  
\begin{equation}
    D_\textrm{FoG}^P(k,\mu,\sigma_\textrm{FoG}^P)=\frac{1}{\left(1+k^2\mu^2\sigma_P^2/2\right)^2},
\end{equation}
where $\sigma_P$  is a free parameter to be constrained by the data.

The SPT redshift space bispectrum can then be written in the following way at tree-level order,
\begin{equation}
B^{\rm SPT}(\textbf{k}_1,\textbf{k}_2,\textbf{k}_3)=D_\textrm{FoG}^B(\textbf{k}_1,\textbf{k}_2,\textbf{k}_3)\left[2Z_1^\textrm{SPT}(\textbf{k}_1)Z_1^\textrm{SPT}(\textbf{k}_2)Z_2^\textrm{SPT}(\textbf{k}_1,\textbf{k}_2)P^L(k_1)P^L(k_2) + \textrm{2perm.}\right],
\label{eq:bispredshiftsp}
\end{equation}
with the kernels $Z_1^\textrm{SPT},Z_2^\textrm{SPT}$ being
\begin{align}
    Z_1^\textrm{SPT}(\textbf{k})&=b_1+f\mu^2,\nonumber\\
    Z_2^\textrm{SPT}(\textbf{k}_1,\textbf{k}_2)&=b_1F_2^\textrm{SPT}(\textbf{k}_1,\textbf{k}_2)+f\mu_{12}^2G_2^\textrm{SPT}(\textbf{k}_1,\textbf{k}_2)+\frac{b_1f}{2}\left(\mu_1^2+\mu_2^2+\mu_1\mu_2\left(\frac{k_1}{k_2}+\frac{k_2}{k_1}\right)\right)\nonumber\\
    &+f^2\mu_1\mu_2\left(\mu_1\mu_2+\frac{1}{2}\left(\mu_1^2\frac{k_1}{k_2}+\mu_2^2\frac{k_2}{k_1}\right)\right)+\frac{1}{2}\left(b_2+b_{s^2}S_2^\textrm{SPT}(\textbf{k}_1,\textbf{k}_2)\right),\label{eq: z1z2}
\end{align}
where $\mu_{ij}\equiv(k_i\mu_i+k_j\mu_j)/|\textbf{k}_i+\textbf{k}_j|$. The $G_2^\textrm{SPT}$ and $S_2^\textrm{SPT}$ kernels in SPT are 
\begin{align}
    G_2^\textrm{SPT}(\textbf{k}_1,\textbf{k}_2)&=\frac{3}{7}+\frac{1}{2}\cos(\theta_{12})\left(\frac{k_1}{k_2}+\frac{k_2}{k_1}\right)+\frac{4}{7}\cos^2(\theta_{12}),\\
    S_2^\textrm{SPT}(\textbf{k}_1,\textbf{k}_2)&=\cos(\theta_{12})^2-\frac{1}{3}.
\end{align}
For the bispectrum, we parametrize the FoG damping factor as \cite{Scoccimarro:1999ed,Verde1998}
\begin{equation}
\label{eq:fog_lorentz}
    D_\textrm{FoG}^B(\textbf{k}_1,\textbf{k}_2,\textbf{k}_3)=(1+\left[k_1^2\mu_1^2+k_2^2\mu_2^2+k_3^2\mu_3^2\right]^2\sigma_{B}^4/2)^{-2},
\end{equation}
again with $\sigma_B$ being a free parameter.

In a realistic application, where objects angular positions and redshifts are transformed into co-moving coordinates assuming a fiducial cosmology, the Alcock-Paczyński effect on the survey's summary statistics  needs to be accounted for \cite{Alcock:1979mp}.

For the power spectrum and bispectrum this is usually done by introducing  the line-of-sight and plane-of-the-sky dilation parameters $\alpha_\parallel, \alpha_\bot$,  which modify $k$ and $\mu$  as \cite{ballinger1996measuring,beutler2014clustering,gil-marin_clustering_2017},
\begin{align}
    k\to \frac{k}{\alpha_\bot}\left[1+\mu^2\left[\left(\frac{\alpha_\bot}{\alpha_\parallel}\right)^2-1\right]\right]^{1/2};\quad \mu\to \mu\frac{\alpha_\bot}{\alpha_\parallel}\left[1+\mu^2\left[\left(\frac{\alpha_\bot}{\alpha_\parallel}\right)^2-1\right]\right]^{-1/2}.
\end{align}

We model the deviations from Poissonian shot-noise with the parameters $A_\textrm{P},A_\textrm{B}$, which modify the Poisson prediction as in \cite{Gil-Marin:2014pva,gil-marin_clustering_2017}:
\begin{align}
    P_{\textrm{noise}}&=(1-\frac{A_\textrm{P}}{\alpha_\parallel \alpha_\bot})P_\textrm{Poisson},\\
    B_{\textrm{noise}}(k_1,k_2,k_3)&=(1-\frac{A_\textrm{B}}{\alpha_\parallel^2\alpha_\bot^4})B_\textrm{Poisson}(k_1,k_2,k_3).
\end{align}

The power spectrum and bispectrum redshift space multipoles are then obtained by integrating the expansion of the power spectrum and bispectrum dependence on the angle with respect to the line of sight  in terms of Legendre polynomials $\mathcal{L}_i$, so that
\begin{align}
    P^{(\ell)}(k)&=\frac{2\ell+1}{2\alpha_\parallel\alpha_\bot}\int_{-1}^1d\mu P(k,\mu)\mathcal{L}_{\ell}(\mu),\label{eq: pmulti}\\
    B^{(\ell_i)}(\textbf{k}_1,\textbf{k}_2,\textbf{k}_3)&=\frac{2\ell+1}{4\pi\alpha_\parallel^2\alpha_\bot^4}\int_{-1}^1d\mu_1\int_0^{2\pi}d\phi B(\textbf{k}_1,\textbf{k}_2,\textbf{k}_3)\mathcal{L}_{\ell}(\mu_i),
\end{align}
Here $\phi$ is defined as the angle fulfilling $\mu_2=\mu_1\cos\theta_{12}-\sqrt{(1-\mu_1^2)(1-\cos\theta_{12}^2)}\cos\phi$, and $\ell_i$ refers to the multipole of order $\ell$ ($\ell=0,2$ corresponding respectively to the monopole and quadrupole), and the $i$ index denotes which multipole (for instance $\ell=2,$ $\ell_1$ being the quadrupole $(200)$). The power spectrum multipole expansion of \ref{eq: pmulti} was proposed in \cite{hamilton1992measuring,cole1994fourier}, while the bispectrum expansion and choice of variables was first used in \cite{Scoccimarro:1999ed}. %
Note that the expansion about the LOS of the bispectrum is not unique; %
for example in \cite{sugiyama2019complete} %
the bispectrum anisotropic signal is expanded in terms of the spherical harmonic decomposition with zero total angular momentum.

The tree-level SPT model for the bispectrum is only valid on large scales, typically for $k\lesssim 0.07\,h{\rm Mpc}^{-1}$ at $z\sim0.5$. Note that this $k$-range is much more restrictive than that routinely used for the power spectrum (with 1-loop or 2-loop SPT modelling) which extends typically to 
$k\sim 0.15\,h{\rm Mpc}^{-1}$. 

A bispectrum modeling  beyond the tree-level formalism is needed 
to  match the $k$-range of applicability for both power spectrum and bispectrum without introducing  unacceptable biases. %
A valid approach is to perform loop corrections on the bispectrum, see e.g.,  \cite{Sefusatti:2009qh,philcox2022cosmology,eggemeier2021testing}.

Alternatively, 
inspired by previous works \cite{Scoccimarro2001,GilMarin:2011ik,Gil-Marin:2014pva}, we propose a phenomenologically motivated bispectrum model, based on tree-level SPT, with few  coefficients  which we calibrate on simulations, and demonstrate that it is robust against variation in redshift, tracer-type and cosmology.

\begin{figure}[ht]
\centering
\hspace*{1cm}\includegraphics[ width = 0.82\textwidth ]
{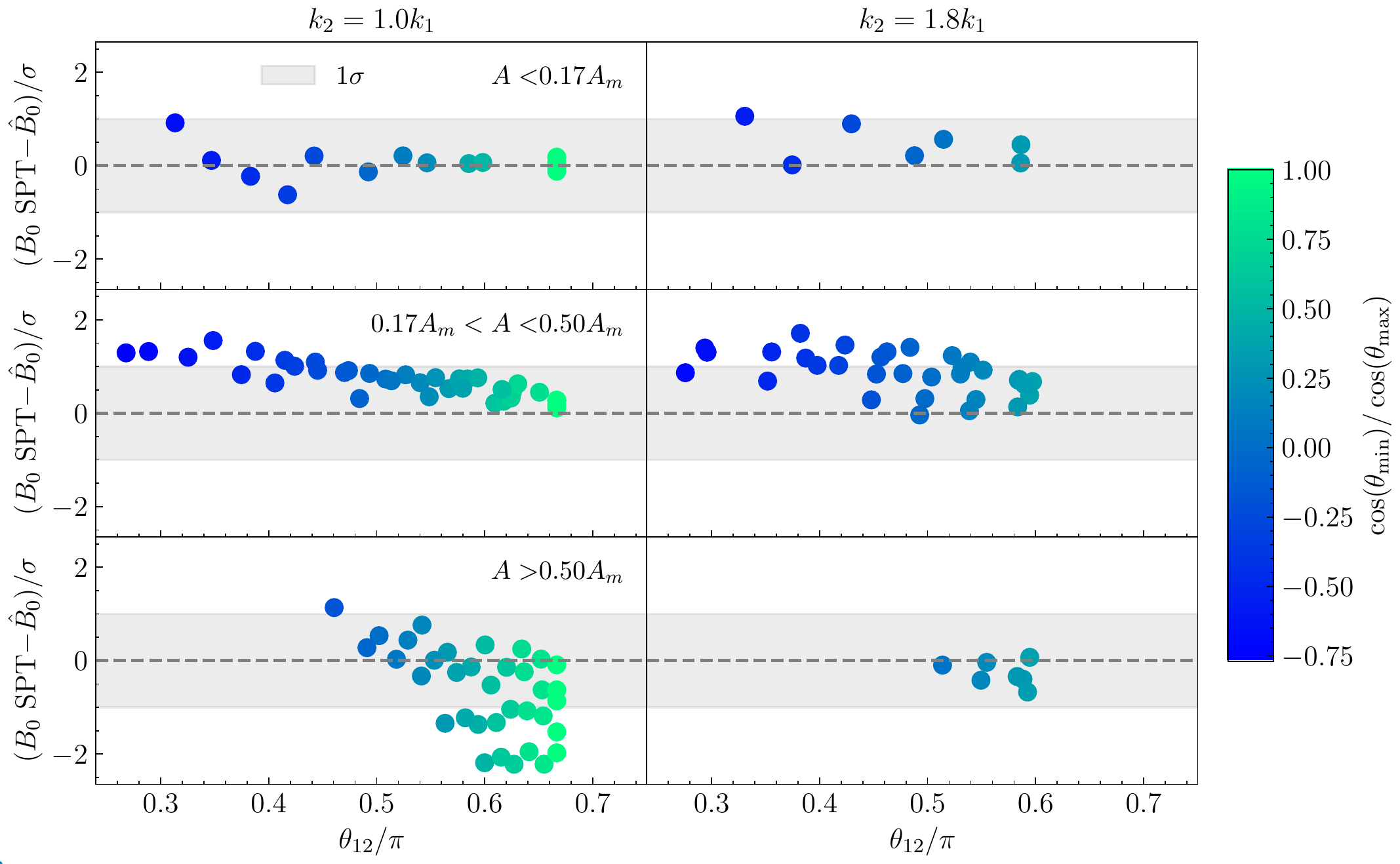}
\hspace*{-0.67cm}\includegraphics[ width = 0.702\textwidth ]
{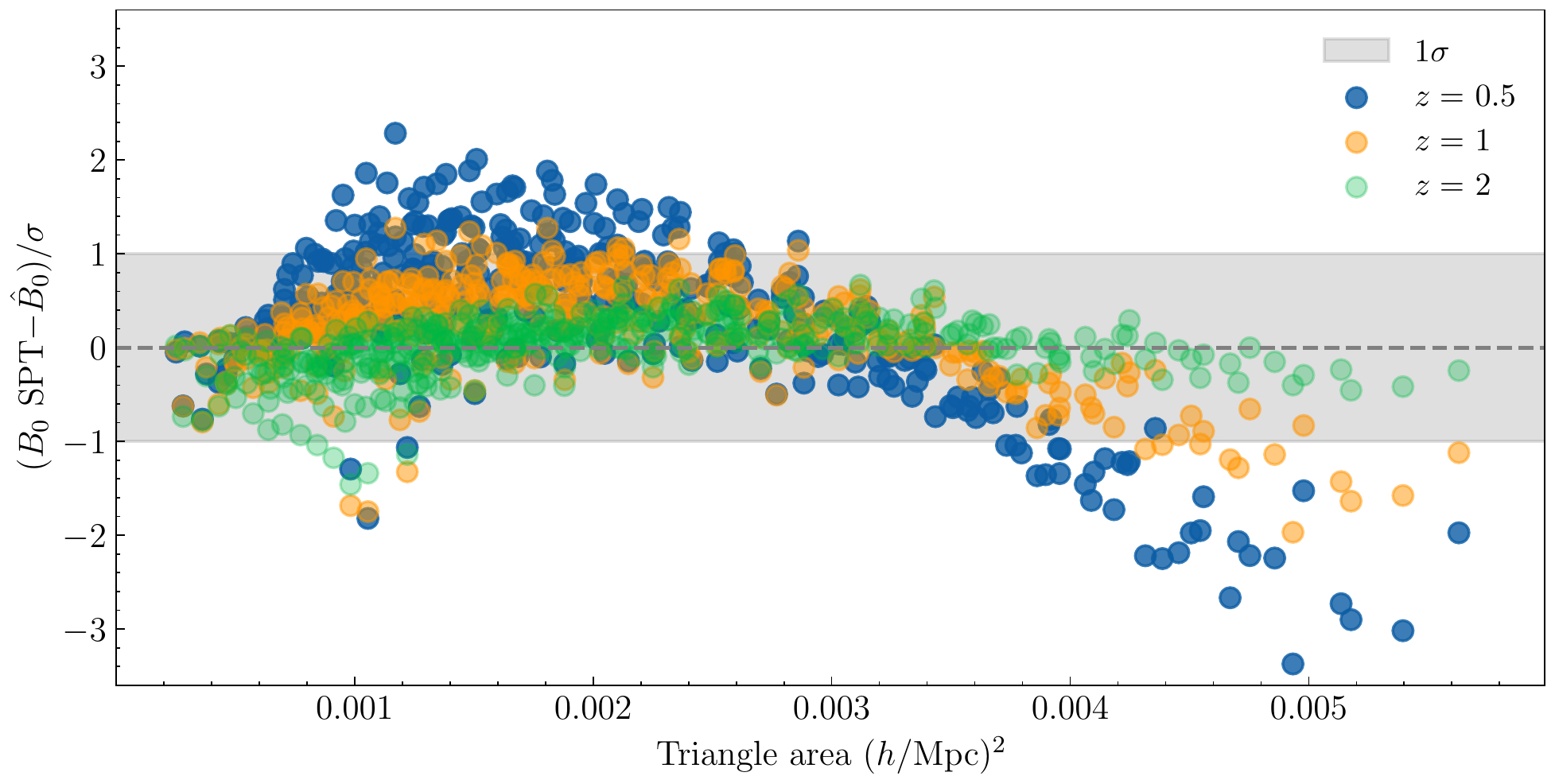}

\caption{Normalized residuals for the best fit of the SPT redshift-space bispectrum, as a function of the shape and area of the triangles, for $0.02<k<0.12$. The only free parameter in the fit is $\sigma_B$, governing the incoherent velocity dispersion (FoG) effect.  In the panel above we show, for two  representative values of $k_2/k_1$ and for $z=0.5$, the residuals of the SPT bispectrum as a function of the angle $\theta_{12}$ between the vectors $\textbf{k}_1$ and $\textbf{k}_2$. The three rows correspond to triangles with different areas, expressed as fraction to the maximum area, $A_m$. The normalization $\sigma$ is the expected  standard deviation corresponding to a volume of $100$ $ (\textrm{Gpc}\,h^{-1})^3$.
The colourbar shows the quantity $\cos(\theta_\textrm{max})/\cos(\theta_\textrm{min})$, which is highly sensitive to the shape of the triangle. %
The lower panel shows the direct dependence between area and accuracy of SPT bispectrum, for $z=0.5,1,2$.
}
\label{fig:B0_Quijfid_red_spt}
\end{figure}

We build on  the findings of \cite{gualdi_geometrical_2019,Gualdi:2019sfc}, where the information content of bispectrum is organized by geometric coordinates of the triangles,  and investigate to what extent the bispectrum model can be improved by taking into account these coordinates. To gain physical intuition, %
Figure \ref{fig:B0_Quijfid_red_spt} illustrates which features of the gravitational bispectrum signal the tree-level SPT model fails to capture. The figure relates the residuals of the  tree-level SPT bispectrum 
normalized by the statistical errorbars of the simulations (normalized residuals), to specific geometric quantities of the triangles. 
In this figure, and in the rest of the paper, we impose
$|\textbf{k}_1|\leq|\textbf{k}_2|\leq|\textbf{k}_3|,$ with the specific choices of $k$-range and $\Delta k$ being presented in Section~\ref{sec: methodology}. In the upper panel of Fig.~\ref{fig:B0_Quijfid_red_spt}, we display two subsets of triangles, with the ratios $k_2/k_1=1.0,\, 1.8$, which we divide in three bins of area $A$\footnote{The area of a triangle can easily be computed given the absolute values of $k_1,k_2,k_3$ by Heron's formula: $A=\sqrt{s(s-k_1)(s-k_2)(s-k_3)}$, $s$ being the semiperimeter of the triangle, $s=\frac{1}{2}(k_1+k_2+k_3)$.}, represented by the 3 sub-panels, $A<0.17A_m$, $0.17A_m<A<0.50A_m$, $A>0.50A_m$, where $A_m$ denotes the maximum area. In each of these 6 binned subsets of triangles, the normalized residuals are shown as function of the exterior angle $\theta_{12}$ between $\textbf{k}_1$ and $\textbf{k}_2.$ We colour each data-point according to its value of the ratio $\cos(\theta_{\max})/\cos(\theta_{\min})$, where $\theta_{\max},\theta_{\min}$ are the maximum and minimum (exterior) angles of the triangle. In the lower panel, we show the relationship between the area of triangles and their associated SPT residual.
 
It is expected that the accuracy of the  SPT bispectrum model degrades as $k_1$ increases. In this set up, for a fixed $\theta_{12}$ and a fixed $k_1/k_2$ ratio, the area is determined by $|k_1|$ and the SPT bispectrum accuracy shows a trend with the area, where larger area triangles are less accurately modeled. There is also a monotonic  dependence on  $\theta_{12}$. As noted already in the literature \cite{scoccimarro1998nonlinear,smith2008analytic},  for a given $k_2$-to-$k_1$ ratio and a given scale, the  accuracy of the SPT bispectrum model depends on the triangle shape i.e., on  $\cos{\theta_{12}}$ (as illustrated by the color scheme).

The dependence on the area is summarized in the bottom panel of Figure~\ref{fig:B0_Quijfid_red_spt}: there is a
clear dependence  of the  bispectrum residuals with the  area of triangles,
that becomes more marked with the growth of structure at lower redshifts. Several triangle shapes (several $k_1/k_2$ ratios and several angles $\theta_{12}$) have a given area, and the residual dependence  on these quantities gives rise to a lot of the scatter.

\subsection{Proposed model: Geometric FPT (GEO-FPT) model}
\label{sub:GEO}
In what follows we present the Geometric Fitted Perturbation Theory model, which we will refer to as GEO-FPT (GEO in superscripts), that %
models the dependence of the bispectrum with the geometric properties of the triangles. 
Motivated by the findings shown in Figure \ref{fig:B0_Quijfid_red_spt},
we  begin by including  the dependence on the triangle area $A$, by adding two extra terms to the $Z_2^\textrm{SPT}$ kernel, respectively proportional to $A/A_\textrm{norm}$ and $(A/A_\textrm{norm})^2$ --$A_\textrm{norm}$ is a constant normalization factor which we define to be $A_\textrm{norm}=0.001$ $(h\textrm{ Mpc}^{-1})^2$. Two additional corrections tackle the dependence on angle. These terms are built from a  combination of ratios of cosines of $\theta_\textrm{max},\theta_\textrm{med},\theta_\textrm{min}$ (respectively the maximum, intermediate and minimum external angles of the triangle). The first correction, $\cos(\theta_\textrm{med})/\cos(\theta_\textrm{max})$, quantifies how squeezed the triangle is, while the second ratio, $\cos(\theta_\textrm{min})/\cos(\theta_\textrm{max})$, quantifies how flattened it is\footnote{For triangles satisfying $k_1\leq k_2\leq k_3$, a triangle will be considered to be squeezed if $k_3\sim k_2\gg k_1$, and flattened if $k_1\sim k_2\gtrsim k_3/2$.}. Finally, we add a constant parameter governing the overall amplitude of the bispectrum signal. In summary, we transform $Z_2^\textrm{SPT}\to Z_2^\textrm{GEO},$ as follows

\begin{equation}\label{eq: Z2_geo}
    Z_2^{\textrm{GEO}} = Z_2^\textrm{SPT}\times\Big[f_1+f_2\frac{\cos(\theta_\textrm{med})}{\cos(\theta_\textrm{max})}+f_3\frac{\cos(\theta_\textrm{min})}{\cos(\theta_\textrm{max})}+f_4\frac{A}{A_\textrm{norm}}+f_5\frac{A^2}{A_\textrm{norm}^2}\Big].
\end{equation}
where the coefficients $f_1,...,f_5$  should be calibrated on N-body simulations and mock galaxy surveys.
In this model,  the $F_2$ and $G_2$ kernels maintain the same expressions as in SPT and the correction is the same for all  redhsift space multipoles.  
Besides the $Z_2$ kernel, the other change of the GEO-FPT model, compared to Equations~\ref{eq: bisp_spt} and \ref{eq:bispredshiftsp},  is that the input power spectrum  is  not the linear one $P^L$ but rather the non-linear power spectrum $P_{\delta\delta}$ from Equation ~\ref{eq: Pnl_matter}. In Appendix \ref{App: Nseries} we  explore whether having different best-fit parameters $f_1,...,f_5$  for the monopole and quadrupoles can yield an improvement in accuracy and precision, finding no substantial gain in doing so. 

The FoG parameters $\sigma_P,\sigma_B$   are  always  treated as  nuisance parameters and marginalized over.
The shot-noise parameters $A_\textrm{P},A_\textrm{B}$  are kept fixed at their fiducial values of unity when fitting the $\{ f_1,..., f_5\}$ kernel coefficients, then they are used as nuisance parameters in  parameter  inference  and thus marginalized over.

In summary, the complete form of the GEO-FPT bispectrum reads as 
\begin{align}
B^{\rm GEO}(\textbf{k}_1,\textbf{k}_2)=D_\textrm{FoG}^B(\textbf{k}_1,\textbf{k}_2)\left[2Z_1^\textrm{SPT}(\textbf{k}_1)Z_1^\textrm{SPT}(\textbf{k}_2)Z_2^\textrm{GEO}(\textbf{k}_1,\textbf{k}_2)P^{\rm NL}(k_1)P^{\rm NL}(k_2)+ \textrm{2perm.}\right] ,
\label{eq:bispredshiftspGEO}
\end{align}
where $P^{\rm NL}$ indicates the real space non-linear power spectrum, and $\textbf{k}_3=-\textbf{k}_1-\textbf{k}_2$ is implicit. In this paper  $P^{\rm NL}$ is given by Equation \ref{eq: Pnl_matter} and computed at  2L-RPT with \textsc{PTcool}\footnote{\url{https://github.com/hectorgil/PTcool}} and \textsc{Brass}\footnote{\url{https://github.com/hectorgil/Brass}}, but other models (such as \textsc{Halofit} \cite{smith2003stable,takahashi2012revising}) could be used.

\section{Methodology}
\label{sec: methodology}

We fit the 5 free parameters of the $Z_2^{\textrm{GEO}}$ Eq.~\ref{eq: Z2_geo} to measurements of bispectrum from a large sample of dark matter N-body simulations provided by the \textsc{Quijote} suite \cite{villaescusa-navarro_quijote_2020} (8000 independent realizations at $z=0.5$ and 4000 realizations at $z=1,2$), where the measurements are carried out using the pipeline of \cite{gualdi_matter_2021,gualdi_joint_2021}.
When calibrating the $f_1,...,f_5$ parameters with the \textsc{Quijote} simulations, the cosmological parameters $\sigma_8,f,\alpha_\parallel,\alpha_\bot$ are kept fixed at the respective true values, together with the bias parameters $b_1,b_2$ and the shot-noise parameters $A_\textrm{P},A_\textrm{B}$. Then the parameters $f_1,...,f_5,\sigma_B$ are left free to vary, with effectively improper uniform priors\footnote{The priors are sufficiently broad that the MCMC never samples the prior boundary.}. On the other hand, when using the bispectrum model to recover the cosmology (from Section \ref{sec: Quijresults}), the $f_1,...,f_5$ kernel parameters are kept fixed at the best-fit value. All other parameters have again broad uniform priors, with the exception of $A_\textrm{P},A_\textrm{B}$, for which we use Gaussian priors\footnote{This prior is physically-motivated: in the \textsc{Quijote} simulations the shot noise is negligible.} with mean 1 and {\it rms} 0.5. This prior becomes unimportant when the shot noise matters (i.e., in the \textsc{Nseries} simulations).%
We also employ other sets of simulations to assess the performance of the GEO-FPT bispectrum model for different cosmologies and for computing the necessary covariance matrices as we describe in the following section. 

\subsection{Synthetic catalogues}
\label{sec: sims}
We use  four independent  sets of simulations, summarized in Table~\ref{tab:sims}.
The baseline  results are presented for the \textsc{Quijote} simulations, which 
 feature $512^3$ dark matter (DM) particles, evolving gravitationally in a periodic box of side $L_{\rm box}=1$ $\textrm{Gpc}\,h^{-1}$, where initial conditions are set by 2LPT (second-order Lagrangian perturbation theory  \cite{scoccimarro1998transients,crocce2006transients,michaux2021accurate}).  We re-use the 8000 power spectrum and bispectrum measurements performed in the previous work \cite{gualdi_joint_2021}, and we further estimate the bispectrum on 4000 realizations for $z=1,2$.  In order to determine the covariance of these measurements, we use the same 4000 (8000) realizations and re-scale the covariance so it corresponds to that of a large-enough volume\footnote{Since our adopted data vector, as explained in Section \ref{subsec:analysissetup}, is the mean of the statistics from the 8000 (4000) realizations, it effectively corresponds to the signal of a volume of 8000 (4000) $({\rm Gpc}\,h^{-1})^3$. } of 100 $({\rm Gpc}\,h^{-1})^3$ . Since the data-vector has of the order of $\sim 1200$ elements per redshift bin,  this choice of 4000 realizations strikes a good compromise between computational time and accuracy of signal and covariance matrix.
 The cumulative volume of the 4000 (8000) simulations is enough to make the statistical errors in the measured bispectrum signal completely negligible compared to the size of the errors given by the covariance we apply: the fitting residuals are entirely due to modeling systematics.

\begin{table}[ht]
\centering
\begin{tabular}{|l||c|c|c|c|c|c|c|c|}
\hline

 Name (type)   &  $\Omega_\textrm{m}$   & $\Omega_\textrm{b}$  & $\sigma_8$ & $n_\textrm{s}$& $f$ &  $h$   & $N_\textrm{sims}$ & $L_\textrm{box}$   \\ \hline\hline
\textsc{Quijote} (DM) & 0.3175 & 0.049 & 0.834 & 0.9624  & 0.534 & 0.6711 & 8000; 4000 & 1\\  \hline
\textsc{Nseries} (Gal.)& 0.286 & 0.047 & 0.82 & 0.96& 0.504  & 0.6711 & 7 & 2.6 \\  \hline
\textsc{Patchy} (Gal.) & 0.307 & 0.048 & 0.829 & 0.961& 0.525 & 0.678 & 2000 & 2.5 \\  \hline
\textsc{CW}-Fid (DM)& 0.27 & 0.047 & 0.79 & 0.95& 0.483 & 0.7 & 160 & 2.4  \\  \hline
\textsc{CW}-0.2 (DM)& 0.2 & 0.035 & 0.872 & 0.95 & 0.413  & 0.8133 & 1 & 2.79\\  \hline
\textsc{CW}-0.4 (DM)& 0.4 & 0.07 & 0.693 & 0.95& 0.604 & 0.5751 & 1 & 1.97  \\  \hline
\textsc{CW}-1 (DM)  & 1 & 0.173 & 0.493 & 0.95& 1  & 0.3637 & 1 &  1.25 \\  \hline
\end{tabular}

\caption{Main cosmological parameters for all sets of simulations considered in this work. The \textsc{Quijote} simulations (the dark matter particles distribution) are  used for calibrating the model at $z=0.5,1.0,2.0$, as well as a first consistency check. The \textsc{Nseries} measurements (galaxy mocks, where galaxies are painted on top of N-body halos) are used for checking that the model recovers the underlying cosmological parameters in a different cosmology and that it works with biased tracers. The \textsc{Patchy} mocks consist of a large set of 2000 realizations of fast-mocks (not N-body) which mimic the 2- and 3-point statistics observed by BOSS. The \textsc{CW}-Fid, 0.2, 0.4, 1 simulations (dark matter particles distribution) allow an assessment of the validity of the model as $\Omega_\textrm{m}$ is varied, while leaving $\Omega_\textrm{m}h^2$ unchanged. $L_\textrm{box}$ is expressed in units of $\left[\textrm{Gpc}\,h^{-1}\right]$.  %
 }\label{tab:sims}
\end{table}

The underlying cosmology of this simulations set is characterized by the following cosmological parameters: the matter and baryon densities, $\Omega_{m}=0.3175,\Omega_b=0.049$, the amplitude of dark matter fluctuations, $\sigma_8=0.834$, the scalar spectral index, $n_s=0.9624$, the reduced Hubble constant, $h=0.6711$, and the neutrino mass $M_\nu=0$; which are fully compatible with Planck best-fit results \cite{collaboration2018planck}.

To assess how the model captures the cosmology and galaxy-bias model dependencies, 
we  employ  the \textsc{Nseries} periodic box mocks\footnote{\url{https://www.ub.edu/bispectrum/page12.html}} \cite{satpathy2017clustering,alam2017clustering,hahn2017effect}. This is a set of 7 independent realizations of dark matter simulations at $z=0.5$, where haloes  are populated with galaxies according to a Halo Occupation Distribution compatible with BOSS clustering data, where the average density of galaxies is $\bar{n}\simeq 4\times 10^{-4}$. The cosmological parameters are consistent with WMAP9 best-fit cosmology \cite{hinshaw2013nine}: $\Omega_\textrm{m}=0.286,\,\Omega_\textrm{b}=0.047,\,\sigma_8=0.82,\,n_\textrm{s}=0.96,\,h=0.7$. Each realization consists of a cubic box whose side is $L_{\rm box}=2.6\,{\rm Gpc}\,h^{-1}$. The total physical volume for the 7 boxes is 123 $(\textrm{Gpc}\,h^{-1})^3$, corresponding to an effective volume of 80 $(\textrm{Gpc}\,h^{-1})^3$, with the effective volume computed from the physical volume following \cite{tegmark1997measuring}. The full data-vector for the \textsc{Nseries} analysis has 441 elements; since 7 realizations are not enough for estimating the covariance of this data-vector, we employ instead 
the 2000 realizations of the cubic \textsc{Patchy} galaxy fast-mocks \cite{kitaura2016clustering} with $z=0.53$ and $L_{\rm box}=2.5$ Gpc $h^{-1}$, which were also produced for their application to the BOSS analysis. Each of these mocks has on average $\sim 185^3$ galaxies, resulting in $\bar{n}\simeq4\cdot 10^{-4}$, and therefore very close to the signal-to-noise ratio of \textsc{Nseries}. In this case, the cosmological parameters are $\Omega_{\textrm{m}}=0.307,\,\Omega_{\textrm{b}} = 0.048,\,\sigma_8=0.829,\, n_{\textrm{s}}=0.961,\,h=0.678$. We average the covariance of these realizations to match the physical volume of the mean of the 7 \textsc{Nseries} simulations. 

The behaviour and validity of our model is also tested on a wider range of cosmological parameters by including the set of dark matter simulations that was used in previous bispectrum works \cite{GilMarin:2011ik,Gil-Marin:2014pva} and presented in  \cite{deputter2012}, which we refer to as `\textsc{CW}'. The \textsc{CW} simulations consist of a set of 160 realizations of $768^3$ DM particles,  with a cosmology consistent with WMAP9 (see summary Table \ref{tab:sims}), together with one realization at each of the three following values of $\Omega_\textrm{m} = 0.2\,,0.4\,,1.0$,  keeping $\Omega_{\rm m}h^2$ and the primordial amplitude of scalar fluctuations, $A_s$, fixed in real space. These three additional simulations have the same initial conditions as the {\it first-index} realization of the fiducial set (with $\Omega_\textrm{m}=0.27$), and therefore their fluctuations due to cosmic variance are very similar, allowing us to compute cosmology variations by only looking at the relative variations across these four cosmologies. 
All \textsc{CW} periodic boxes have the same size in units of Gpc, $L_\textrm{box}$=3.43 Gpc, and their size in units of Gpc$h^{-1}$ varies according to the value of $h$ in each, being $L_{\rm box}=2.4\,{\rm Gpc}\,h^{-1}$ for the fiducial WMAP9 cosmology. The estimation of their covariance matrix, starting from the \textsc{Quijote} simulations set, is described in Appendix \ref{app: cov}.

All the simulations are flat-$\Lambda$CDM, with dark energy %
equation of state parameter $w=-1$, and their specs are summarized in Table \ref{tab:sims}.

\subsection{Analysis set-up}
\label{subsec:analysissetup}
We consider our ``data vectors'' as the summary statistic's averages of {\it all} the available realizations (the simulation boxes). As such, the statistical error of these data vectors is the one corresponding to the total cumulative simulations volumes (Section \ref{sec: sims}) hence negligible, especially for the \textsc{Quijote} simulations. 

The redshift space power spectrum data-vector consists of the power spectrum monopole and quadrupole, $\{P_0,P_2\}$, from $k_{\rm min}=0.02\,{\rm Mpc}\,h^{-1}$ up to a certain $k_{\rm max}$. For simplicity, we will refer to this power spectrum data-vector simply as the power spectrum, or $P_{02}$. The bispectrum data-vector consists of the three following multipoles, the monopole and the $k_1$- and $k_2$-quadrupoles, $\{B_0,B_{200},B_{020}\}$, respectively. We will refer to this data-vector just as bispectrum, $B_0$, $B_{02}$, depending on whether we want to consider the isotropic component only, or the full anisotropic component. 
Note that for the $B_{200}$ and $B_{020}$ quadrupoles the configurations fulfilling $k_1=k_2$ are equal, so they can't be considered in both quadrupoles --otherwise the covariance matrix would be singular. We opt for just accounting them on $B_{200}$. Therefore, the number of elements on $B_{200}$ is slightly larger than on $B_{020}$.  The data vectors are measured using the methodology and pipeline presented in \cite{gualdi_matter_2021,gualdi_joint_2021}.

 We have found  that using the three $k_1$-, $k_2$- $k_3$-based quadrupoles, $\{B_0,B_{200},B_{020},B_{002}\}$  does not significantly improve the cosmology constraints compared to our baseline data-vector choice $\{B_0,B_{200},B_{020}\}$, as also reported by \cite{d2022boss}. Hence, for simplicity we will not consider the $B_{002}$ quadruple in the analysis of this paper. 

We use a modified version of \textsc{Rustico}\footnote{\url{https://github.com/hectorgil/Rustico}} to compute the bispectrum from the simulations, and the GEO-FPT bispectrum  code.\footnote{\url{https://github.com/serginovell/Geo-FPT}}

For our baseline analysis, we  consider a $k$-range for the power spectrum of $0.02<k [h\,\textrm{Mpc}^{-1}]<0.15$ (unless otherwise stated, in some cases we use a $k_{\rm max}^P=0.12$ [$h\,\textrm{Mpc}^{-1}$]), while for the bispectrum we focus on the range $0.02<k [h\,\textrm{Mpc}^{-1}]<0.12$, both when fitting for the $f_1,...,f_5$ model parameters, and when inferring the cosmological parameters. This $k$-range is chosen as to capture the statistics in the linear and mildly non-linear regime. Furthermore, we find this set-up to be sufficient for representing most triangle configurations of interest, including 
squeezed triangles up to %
$k_3\sim k_2 \sim 5k_1$. 

Special attention must be given to the power spectrum and bispectrum $k$-bins, $\Delta k$. Our choice of  $\Delta k$ results from a trade off between two competing effects. To minimize binning effects,  $\Delta k$ should be as close  as possible to  $k_{\rm f}$. However, a small bin size  yields a large number of triangles, which are highly correlated and for which the covariance matrix should be estimated from simulations. Given the  available number of simulations 
(8000 (4000) realizations for the \textsc{Quijote} simulations), we choose $\Delta k=1.1k_\textrm{f}^\textrm{Q}\approx0.0069\, h \,{\rm Mpc}^{-1}$ for all the analyses involving the \textsc{Quijote} simulations (Sections \ref{sec: results}, \ref{sec: Quijresults}). Also, as commented in Appendix \ref{app: cov}, we use the \textsc{Quijote} covariance for the analysis of the CW simulations in Section \ref{sec: CW}, so we choose $\Delta k=1.1k_\textrm{f}^\textrm{Q}$ again for this case. For the \textsc{Nseries} analysis, we opt for  $\Delta k=0.01\,h\,{\rm Mpc}^{-1}$, motivated by the available number of mocks, 2000.

 As pointed out in \cite{Sefusatti_binning,Oddo_2020,brieden2021shapefit},  there is a so-called binning effect that, if not accounted for, can introduce systematic errors. The wider the bin, with respect to the fundamental frequency $k_\textrm{f}=2\pi/L_\textrm{box}$, the bigger the binning effect.  In fact, the elements of the data vector are binned measurements:  the measured power spectra (resp. bispectra) is an average of all $k$-vectors (triangles) that are contained in the $k$-bin. The theoretical model on the other hand is usually computed at a single representative $k$ per bin. This ``effective'' $k$ is computed for the power spectrum  as  

 \begin{equation}
    k^\textrm{eff}=\frac{1}{N(k)}\sum_{\textbf{q}_i\in k_1}q_i,
\label{eq: keffective_P}
\end{equation}
with $N(k)$ the number of configurations in the bin. For the bispectrum, it translates as
\begin{equation}
    k_i^\textrm{eff}(k_1,k_2,k_3)=\frac{1}{N(k_1,k_2,k_3)}\sum_{\textbf{q}_1\in k_1}\sum_{\textbf{q}_2\in k_2}\sum_{\textbf{q}_3\in k_3}q_i\delta_K(\textbf{q}_1+\textbf{q}_2+\textbf{q}_3),
\label{eq: keffective_B}
\end{equation}
where $i=1,2,3$, and again $N(k_1,k_2,k_3)$ is the number of configurations in the bin.
As stated above, to minimize the binning effect we consider a bin size of $\Delta k=1.1k_\textrm{f}$.

For the power spectrum this procedure  still leaves some residual  effect due to incomplete mode-sampling for the wider bins at low $|\textbf{k}|$. We correct for this  as 
in \cite{brieden2021shapefit} by introducing $G$,  the binning factor, defined as the ratio between the average of the fiducial  model power spectrum of all the $k$ in a given bin and the fiducial model power spectrum evaluated at the effective $k$ for that $k$-bin, 
\begin{equation}
    G(k^{\rm eff}_j)\equiv \frac{\langle P^{\textrm{model}}(k \in {\rm bin}\, j)\rangle}{P^{\textrm{model}}(k^{\rm ef}_j)},
\end{equation}
 in order to correct  the power spectrum measurements as $P^\textrm{meas}\to G^{-1}P^\textrm{meas}$. We have further checked for the effect of the uneven distribution of $\mu$ across all $k$-vectors in a bin, but found its magnitude to be negligible.

In the bispectrum this effect is, in principle, even more   important, since the condition of triangle closure causes each bin to have proportionally less configurations per data vector element  than in the power spectrum. 
Nevertheless, 
the %
nature of our proposed effective formula, the procedure for fitting its coefficients, together with having a small  $\Delta k$ and working  with $k^\textrm{eff}$ defined as  in Equation \ref{eq: keffective_B},  accounts for  most of the effect.
 Still, flattened triangles, which  are most affected by the binning effect \cite{eggemeier2021testing},
are omitted in 
our analysis by imposing the condition 
$k_3-(k_2+k_1)>\Delta k.$ (recall that $k_1\le k_2\le k_3$).

\subsection{Fitting and covariance matrix estimation}
\label{sec:covmat}

The procedure we choose to fit the free parameters of the kernels to the measurements and then to perform parameter inference with our  calibrated GEO-FPT bispectrum model,  is a Markov chain Monte Carlo sampling, MCMC, specifically using the \textsc{emcee} algorithm \cite{Foreman_Mackey_2013}. In the MCMC sampling, we specify broad, uniform priors\footnote{Effectively these are uniform improper priors because the MCMC never samples the prior boundary.}   in the parameters, and calculate the logarithm of the likelihood assuming Gaussianity and applying the Sellentin-Heavens correction \cite{Sellentin:2015waz}, as

\begin{align}
\label{eq: likeli}
\log L\propto -\frac{n}{2}\ln\left(1+\frac{(\textbf{D}_\textrm{meas.}-\textbf{D}_\textrm{th.})\textrm{Cov}_{\textbf{D}_\textrm{meas.}}^{-1}(\textbf{D}_\textrm{meas.}-\textbf{D}_\textrm{th.})^T}{n-1}\right),
\end{align} 
where $n$ is the number of simulations and $\textbf{D}$ denotes the data vector. For fitting the GEO-FPT bispectrum  coefficients $f_1,...,f_5$, the data vector is $\textbf{D}= \{B_0,B_{200},B_{020}\}$, while for parameter inference the  data vector is the full set $\{P_0,P_{2}, B_0,B_{200},B_{020}\}$. The covariance matrix for the bispectrum-only data vector is a subset of the full $P_{02}+B_{02}$ covariance. 

The full covariance matrix for the data vector is thus an essential component  of the analysis. 
The covariance between different bispectra configurations, and the covariance between power spectrum and bispectrum should not be ignored \cite{oddo2021cosmological,chan2017assessment,sato2013impact,colavincenzo2019comparing}. 
 Thus, an analytical computation of the covariance matrix is highly non-trivial:  if possible, the covariance matrix $\textrm{Cov}_{\textbf{D}_\textrm{meas.}}$ should be estimated from simulations. If the required simulations are not available, it is possible to resort to hybrid approaches where the covariance model is calibrated on mocks \cite{Gualdi:2020ymf,biagetti2022covariance,gualdi_matter_2021}. A  detailed account of our  covariance matrix estimation can be found in Appendix \ref{app: cov}.

 Each covariance matrix is estimated from the simulations measurements at each redshift; it is then  rescaled  to obtain errorbars matching the target volume, with specific rescaling factors in Appendix \ref{app: cov}.
 In summary, to fit the $f_1,..., f_5$ coefficients we use a covariance matrix corresponding to the total volume of the simulations (that is the actual statistical error associated with the mean data vector). To do cosmological inference, when needed,  we rescale the covariance matrix to correspond to a survey volume of $\sim 100 ({\rm Gpc}\,h^{-1})^3$.
 
\begin{figure}[htb]
\centering 
\includegraphics[ width = 0.45\textwidth ]
{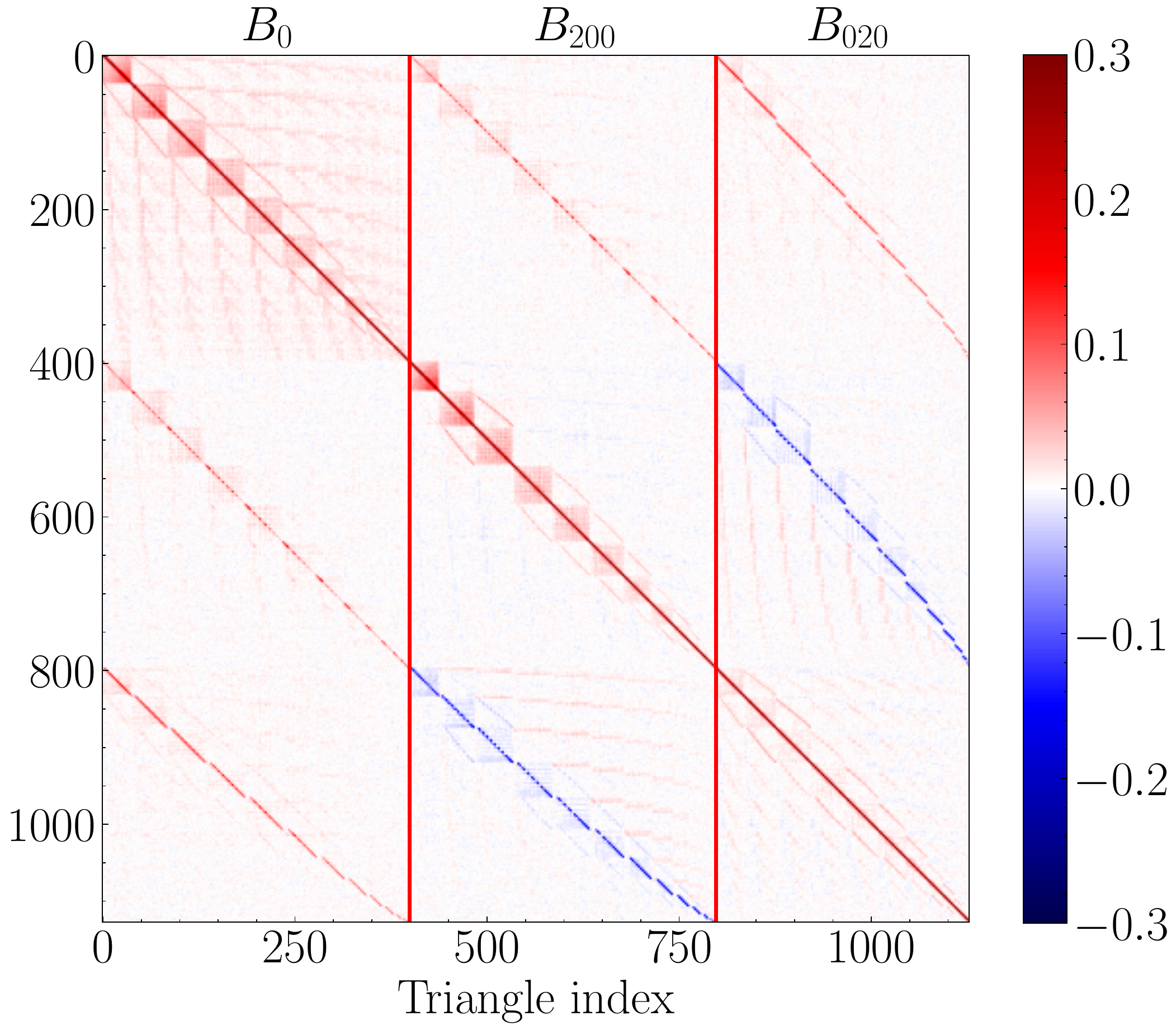}
\includegraphics[width = 0.54\textwidth ]
{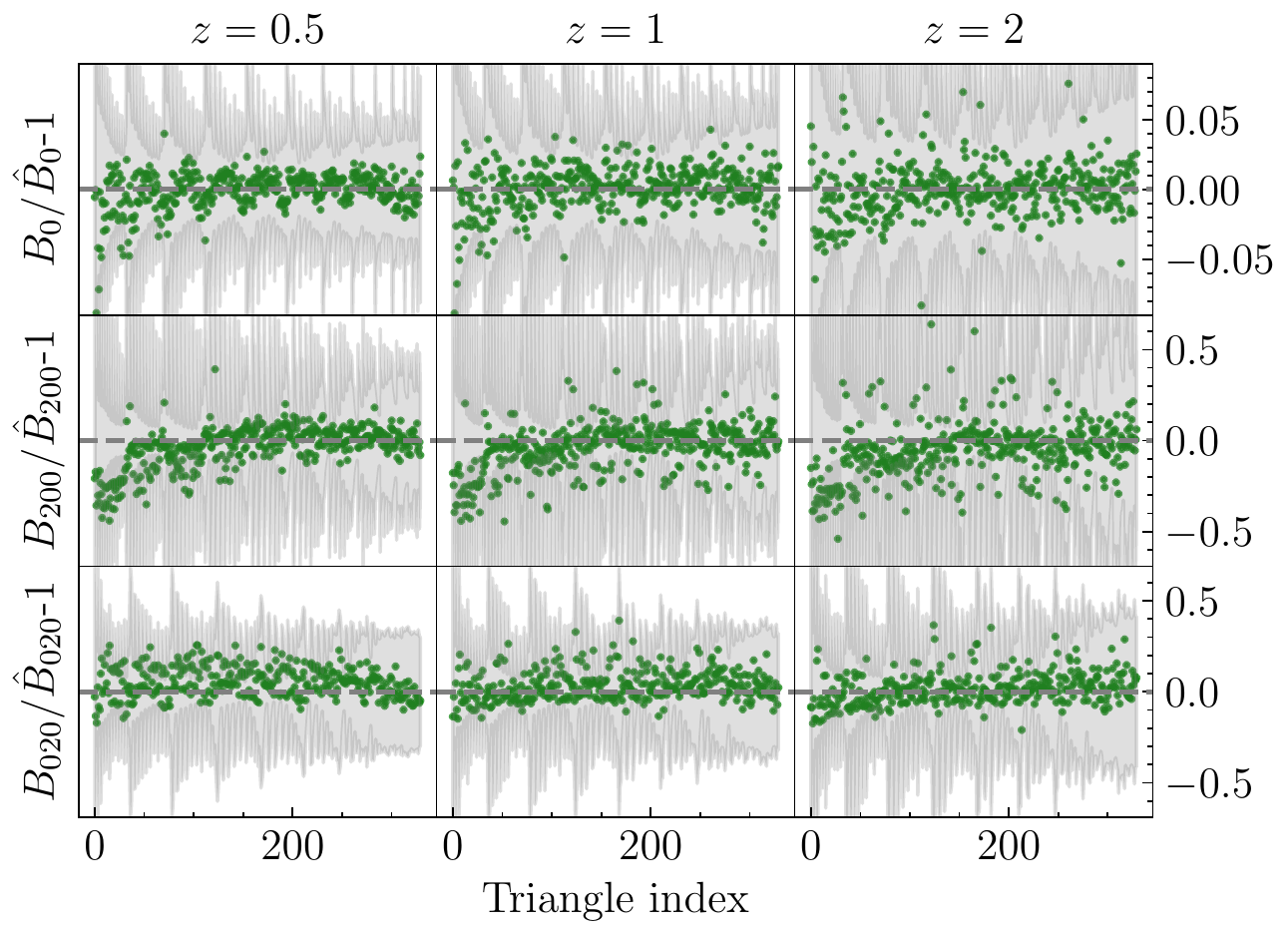}

\caption{Left panel:  reduced covariance matrix of the data-vector formed by $\{B_0,B_{200},B_{020}\}$, obtained with the \textsc{Quijote} simulations at $z=0.5$. The red vertical lines together with the titles in the top x axis indicate how the data vector elements are mapped to the summary statistics. We can observe how, aside from the correlation between the same triangle configuration for different multipoles, there is structure in the covariance matrix, indicating correlation between different configurations. Right panel: ratio between the best fit and the \textsc{Quijote} measurements, for all triangle configurations and all the considered multipoles at each of the snapshots at $z=0.5,1,2$. The cosmology is kept at the fiducial values, and the $k$-range is the same as for the  fit, $0.02<k[h\,\textrm{Mpc}^{-1}]<0.12.$ The dashed grey area represents the variance of the measurements, for a volume of 100 $(\textrm{Gpc}\,h^{-1})^3$. }
\label{fig:B02_Quijfid_red_fit100_sm}
\end{figure}

 The reduced covariance matrix, $R_\textrm{cov}^{ij}=\textrm{Cov}^{ij}/\sqrt{\textrm{Cov}^{ii}\textrm{Cov}^{jj}}$, for the bispectrum-only data vector
 is shown in Figure \ref{fig:B02_Quijfid_red_fit100_sm}, together with the resulting best fit bispectrum for each multipole and redshift. The power spectrum-bispectrum cross-covariance is reported in Appendix \ref{app: cov}.

The best-fit $f_1,...,f_5$ coefficients and their errorbars for each redshift  snapshot can be found in Table \ref{tab: fit_quij}, while the corresponding posterior distributions are shown in Appendix \ref{app: Quijote}, in Figure \ref{fig:matpk_QvCW_comp2LRPT_facc}.

In practice,  we advocate using the estimated best fit values for the GEO-FPT correction coefficients and  interpolating the values of  $\{f_1,f_2,f_3,f_4,f_5\}$ for intermediate redshifts  via spline or linear interpolation, as implemented in \href{https://github.com/serginovell/Geo-FPT}{our code}.

\section{Results}
\label{sec: results}

After the fitting of the coefficients of the $Z_2^{\textrm{GEO}}$ kernel, we obtain a bispectrum theory model whose discrepancy with the measurements from simulations at $z=0.5,1,2$ mostly does not surpass $3\%$ in the case of $B_0$, and $30\%$ in the case of $B_{200}$ and $B_{020}$, as seen in Figure \ref{fig:B02_Quijfid_red_fit100_sm}. This is well within the statistical error for a volume of $\lesssim 100 $ $(\textrm{Gpc}\,h^{-1})^3$. Additionally, we compare the GEO-FPT fit for the monopole at $z=0.5$ against the prescription from SPT in Figure \ref{fig:B0_Quijfid_red_sptfpt} of Appendix \ref{app: Quijote}.

In Figure \ref{fig:Z2kern5par_redCW_mpc100} we show the dependence of the best-fit coefficients on redshift, where the errorbars correspond to the total size of the simulations, 8000 and 4000 $(\textrm{Gpc}\,h^{-1})^3$ for  $z=0.5$ and $z=1,2$ respectively. The shaded region represents the statistical error for a  volume of 100 $(\textrm{Gpc}\,h^{-1})^3$. The dependence with redshift is very mild,  which already suggests  that the $Z_2^{\textrm{GEO}}$ formula is expected to be stable upon changes of the cosmology. We will further show this in Sections \ref{sec: Nseries} and \ref{sec: CW}.

\begin{table}[ht]
\centering
\begin{tabular}{|l||c|c|c|}
\hline
Parameter $(\pm1\sigma)$    &  $z=0.5$               & $z=1$                   & $z=2$\\ \hline \hline
$f_1$     & $1.0033\pm 0.0024$    & $1.0180\pm0.0038$     & $1.037\pm0.0049$          \\ %
$f_2$     & $-0.0040\pm 0.0018$   & $-0.0041\pm0.0031$    & $0.0020\pm0.0041$          \\ %
$f_3$     & $0.0240\pm 0.0014$    & $0.0149\pm0.0024$     & $0.0056\pm0.0033$          \\ %
$f_4$     & $-0.0568\pm 0.0015$   & $-0.0547\pm0.0024$    & $-0.048\pm0.0033$          \\ %
$f_5$     & $0.01325\pm 0.00029$  & $0.01144\pm0.00046$   & $0.0081\pm0.0006$          \\ \hline

\end{tabular}
\caption{Best-fit parameters $\{f_1,...,f_5\}$ for the $Z_2^\textrm{GEO}$ kernel, for the redshifts $z=0.5,1,2$. The fits were performed following the set-up described in \ref{sec: methodology}, using $n_\textrm{sim}=$8000-4000 \textsc{Quijote} simulations respectively for $z=0.5$ and $z=1,2$, for both estimating the signal and the covariance matrix. The covariance matrix is rescaled so that the errorbars correspond to the total volume of the sample, which is equal to $n_\textrm{sim}\times 1 (\textrm{Gpc}\,h^{-1})^3$. }
\label{tab: fit_quij}
\end{table}

\begin{figure}[htb]
\centering 
\includegraphics[width = \textwidth ]
{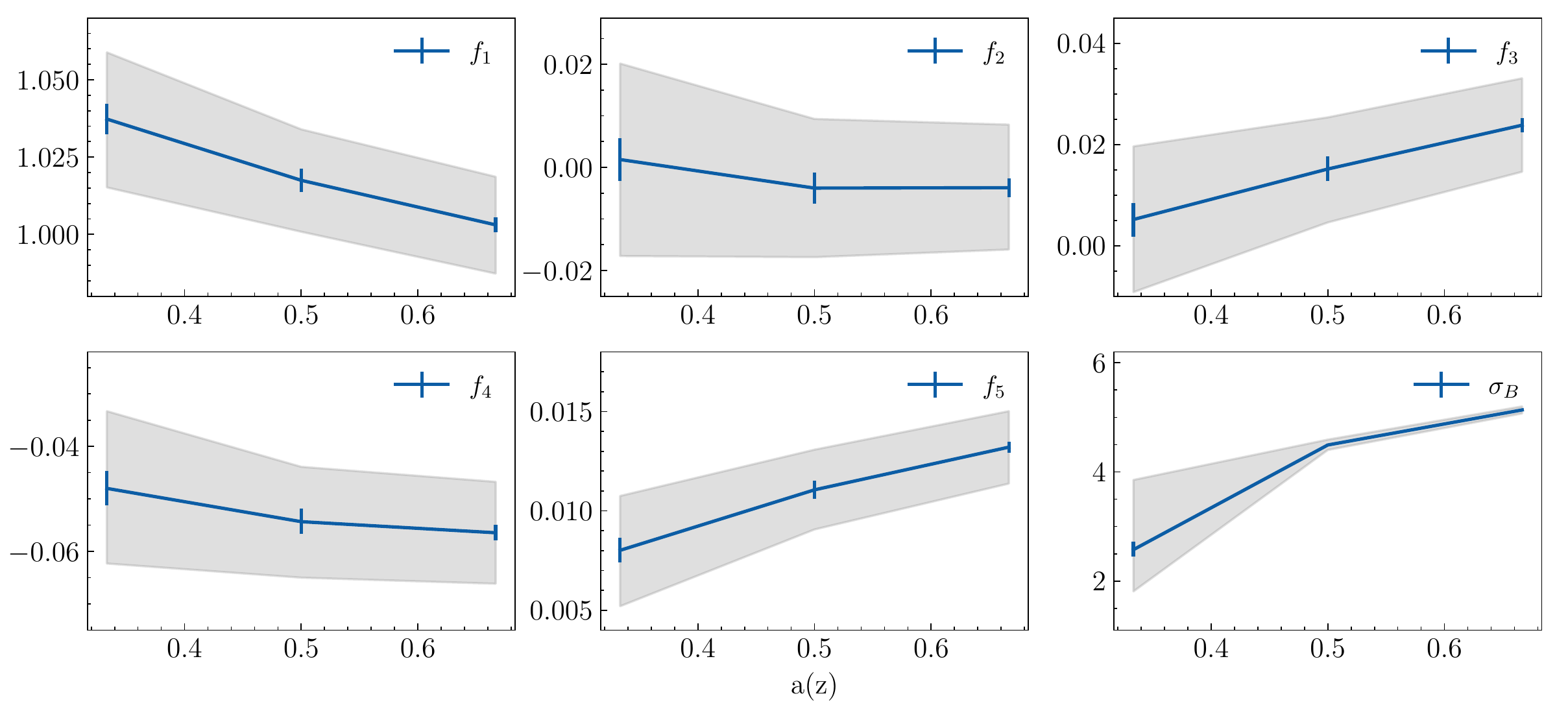}
\caption{Dependence on redshift for the kernel coefficients  $\{f_1,f_2,f_3,f_4,f_5\}$ and the FoG parameter  $\sigma_\textrm{B}$: quantities are shown as a function of   $a(z)=1/(1+z)$ for $z=0.5,1,2$. The best fits have been obtained using the fiducial set of \textsc{Quijote} for both the signal and covariance matrix estimation. We assume no cross-covariance across different redshifts, and at each redshift the covariance used corresponds to the volume of  the total number of simulations. The shaded region shows the resulting $1\sigma$ statistical uncertainty for a volume of 100 $ (\textrm{Gpc}\,h^{-1})^3$. The coefficients are obtained from a fit to the data-vector $B_0+B_{200}+B_{020}$, with a constant $k$-range across redshift and multipoles of $0.02<k (h\,{\rm Mpc}^{-1})<0.12.$ }
\label{fig:Z2kern5par_redCW_mpc100}
\end{figure}

\begin{figure}[htb]
\centering 
\includegraphics[width = \textwidth ]
{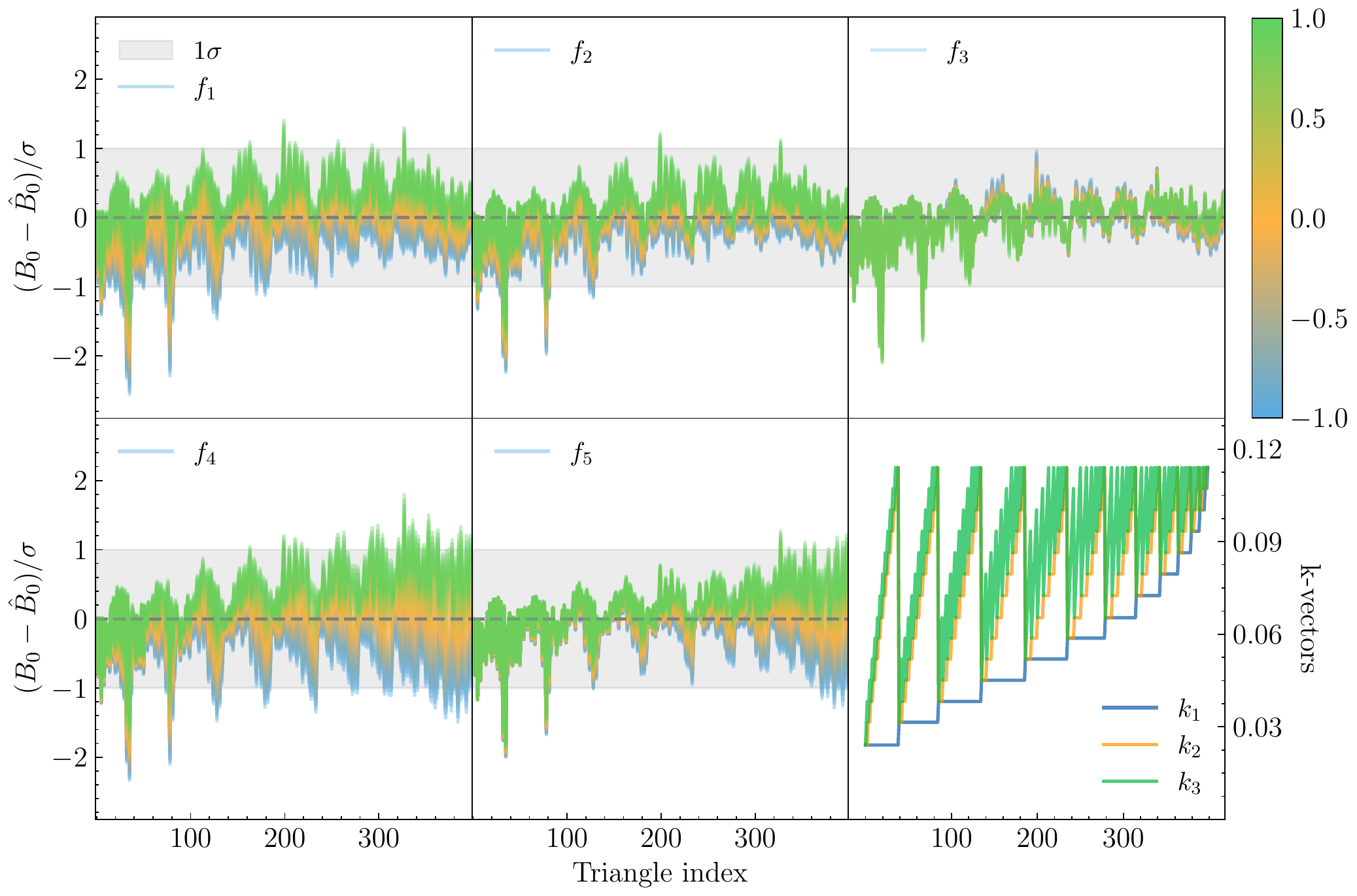}
\caption{We show how (conditional) variations of 1$\sigma$ (described by the colourbar) for each parameter of the $Z_2^{\textrm{GEO}}$ kernel affect the fit of the bispectrum monopole $B_0$ to the mean of the simulations measurements. The bottom right plot shows the correspondence between the configurations and the values of $k_1,k_2,k_3$ of the triangle. }
\label{fig:B0_Quijfid_red_fpt}
\end{figure}

Additionally, in Figure \ref{fig:B0_Quijfid_red_fpt} we  visualize  the effect on the  GEO-FPT bispectrum monopole model of each individual kernel coefficients, at $z=0.5$. In each panel one coefficient is displaced from the best fit by plus/minus one standard deviation, while all other coefficients are kept fixed at their best-fit value.  The bottom-right panel helps with the interpretation of the triangle index in term of triangles shape.

As expected, $f_1$, which acts as an amplitude parameter, uniformly displaces the model across all configurations. The effect of the $f_2$ parameter can be understood as follows: $f_2$ modulates the term 
$\cos\theta_{\textrm{med}}/\cos\theta_{\max}$, which goes to zero for the very squeezed triangles and is 
1 for the equilateral and flattened triangles. Hence $f_2$ leaves unchanged a small number of configurations, while acting very similarly to $f_1$  for all others.
 The $f_3$ term on the other hand  changes sign between flattened and equilateral   configurations; %
  because of this, $f_3$ has a very low correlation with the other parameters (see Figure \ref{fig:matpk_QvCW_comp2LRPT_facc}).
The $f_4$ and $f_5$ parameters, which modulate the terms proportional to $A$ and $A^2$ respectively, obviously become relevant at triangles with large area. We can see the two parameters are complementary since %
changes of $f_5$ target more strongly the non-linear configurations, while $f_4$ affects triangles from the linear regime as well.

\subsection{Performance of the model in recovering cosmological parameters}
\label{sec: results_fs8}
We next quantify how well our  \textsc{Quijote}-calibrated GEO-FPT model  can recover the original parameters of the simulations when used for cosmological inference. Since the statistical errors involved in the fit of the kernel's coefficients are negligible compared to the systematic errors, this procedure tests if the adopted GEO-FPT model is significantly  biased. 

 We test the model first with the \textsc{Quijote} suite of simulations in Section \ref{sec: Quijresults}, to then assess the performance of the model in the case of biased tracers and different cosmology, in Section \ref{sec: Nseries}. In these first two subsections, we focus on the set of parameters $\{f\sigma_8,\alpha_\parallel,\alpha_\bot\}$. In Section \ref{sec: fs8 disentangle} we  focus on the bispectrum  role in breaking 
 the $f\sigma_8$ degeneracy that is naturally present in the power spectrum. Finally, we will provide a further test on unbiased tracers but a cosmology with very different matter density parameter (Section \ref{sec: CW}, \textsc{CW} simulations).
\subsubsection{\textsc{Quijote} simulations}
\label{sec: Quijresults}

\begin{figure}[htb]
\centering 
\includegraphics[  width = \textwidth ]
{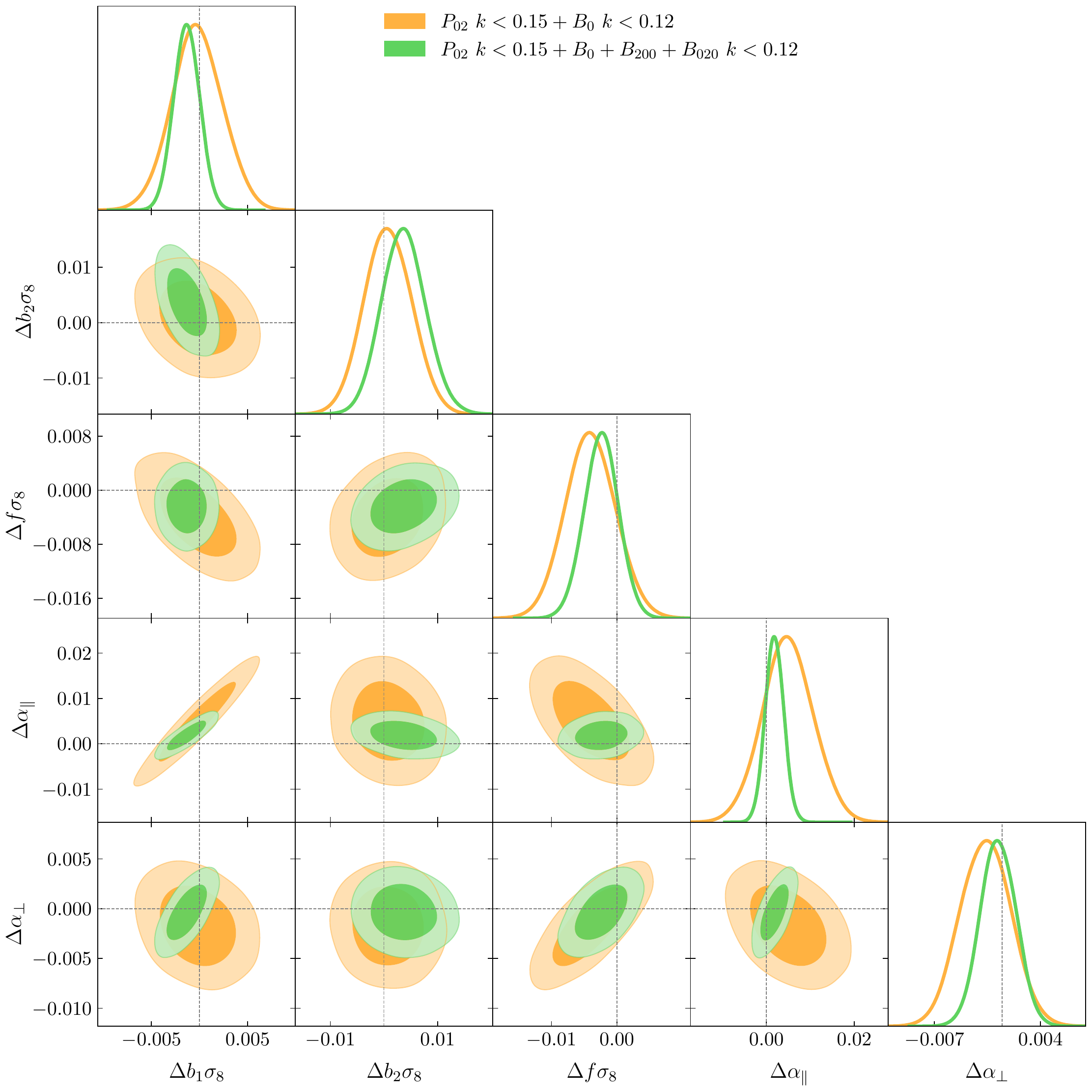}
\caption{Main cosmological parameters recovered from the \textsc{Quijote} fiducial dark matter simulations, using the  (\textsc{Quijote})-calibrated effective $Z_2^{\textrm{GEO}}$ kernel. The parameters are shown in terms of the deviation $\Delta$ from the fiducial (input) value. We show the cases where we use the combination of the power spectrum monopole and quadrupole (obtained as mentioned in \ref{sec: theory}) together with, respectively, $B_0$ and $B_0+B_{200}+B_{020}$. The covariance is obtained from the simulations, and is rescaled  to be equivalent to that of a volume of 100 $(\textrm{Gpc}\,h^{-1})^3$. As expected,  using the bispectrum quadrupoles  tightens the constraints  significantly compared to the case with only $B_0.$ The FoG parameters, $\sigma_\textrm{P}$ and $\sigma_\textrm{B}$, together with the deviations from Poissonian shot-noise $A_\textrm{P}$ and $A_\textrm{B}$, are marginalized over.}
\label{fig:Quijote_P02B02_nog2poc_5parnorm}
\end{figure}

We perform an MCMC sampling of the parameter space  $\{b_1\sigma_8,b_2\sigma_8,f\sigma_8,\alpha_\parallel,\alpha_\bot,A_\textrm{P},A_\textrm{B},\sigma_P,\allowbreak \sigma_B\}$, using the effective $Z_2^\textrm{GEO}$ kernel with its coefficients fixed at the  best fit values (reported in Table \ref{tab: fit_quij}). We refer to this as the (\textsc{Quijote})-calibrated $Z_2^\textrm{GEO}$ kernel.
All priors are effectively improper uniform except for $A_P$ and $A_B,$  where we impose a normal distribution $\mathcal{N}$(1,0.5) as recommended in \cite{HGMeboss}.
In Figure \ref{fig:Quijote_P02B02_nog2poc_5parnorm} we show the 1 and 2D posterior distributions for the cosmological parameters of interest  for the data-vectors $P_{02}+B_0$ and $P_{02}+B_{02}$ at $z=0.5$;  $A_\textrm{P},A_\textrm{B},\sigma_P,\sigma_B$ are considered  nuisance parameters and are marginalized over. The errorbars correspond to a volume of 100 $(\textrm{Gpc}\,h^{-1})^3$, which is suitable to assess whether  possible systematic biases in the recovered parameters are significant compared to the statistical errors expected from future surveys. 
The central values and 1$\sigma$ errors for the cosmological parameters are reported in   Table \ref{tab: results}. 
Results for the $P_{02}+B_{02}$ data-vectors at $z=1,2$ can be found in Appendix \ref{app: Quijote}. %
In all cases the $Z_2^\textrm{GEO}$ kernel is calibrated to the full data-vector $B_0+B_{200}+B_{020}$. 

For all data-vectors considered, all  parameters are recovered  within $1\sigma$ %
(the fiducial values  at this redshift are $f\sigma_8=0.489$, while of course $b_1=1,b_2=0,\alpha_\parallel=\alpha_\bot=1$). The $P_{02}+B_{02}$ data-vector notably reduces  the errorbars  compared to those obtained with $P_{02}+B_0$ (or $P_{02}$ alone), by factors of $63$ and $31\%$ for  $\alpha_\parallel$ and $\alpha_\bot$ respectively (see Table~ \ref{tab: results}).

In the next subsection we test the performance of the GEO-FPT bispectrum model on biased tracers and a slightly different cosmology  (Section \ref{sec: Nseries}, \textsc{Nseries} mocks). In Section \ref{sec: fs8 disentangle} we will explore the improvement that the anisotropic bispectrum provides to the inference of the parameters $f$ and $\sigma_8$ separately.

\subsubsection{Performance on biased tracers for a WMAP-compatible cosmology}
\label{sec: Nseries}
We now turn to the \textsc{Nseries} galaxy simulations. These simulations, described in Section \ref{sec: sims}, differ from \textsc{Quijote} in the fiducial cosmological parameters and, most importantly, allow us to test  the performance of the model in describing the galaxy bispectra, when extended to include the bias expansion of Eq. \ref{eq: z1z2}.
The coefficients of the  the $Z_2^\textrm{GEO}$ kernel  remain fixed to the values obtained in the fit with \textsc{Quijote}.

\begin{figure}[htb]
\centering 
\includegraphics[  width = \textwidth ]
{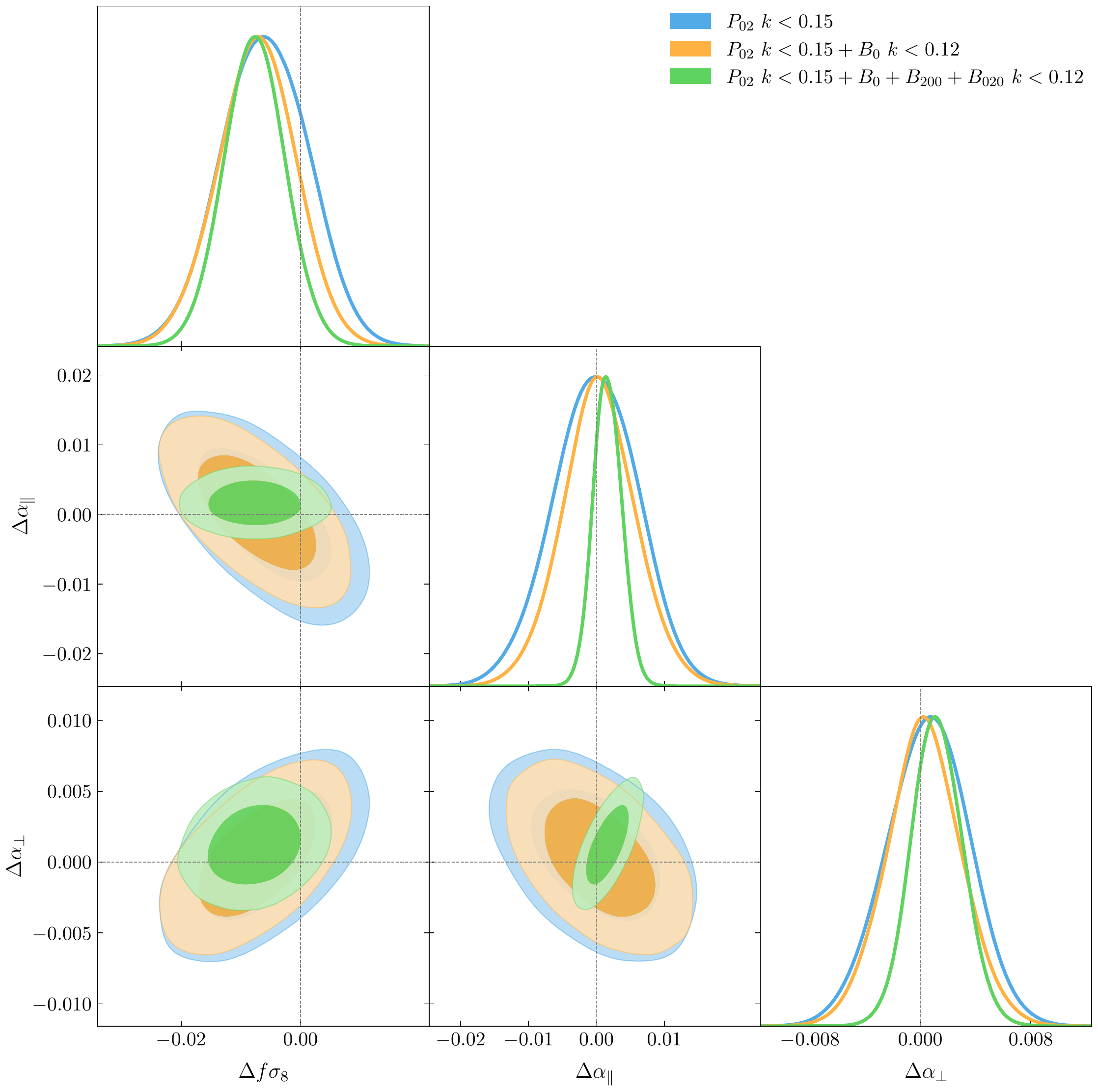}
\caption{Main cosmological parameters recovered from the mean of the seven \textsc{Nseries} halo and galaxy simulations, for the three sequential cases where, respectively, $P_0+P_2$, $P_0+P_2+B_0$ and $P_0+P_2+B_0+B_{200}+B_{020}$ are used. In this case, the covariance is estimated from the Patchy periodic box mocks, and is rescaled %
to match the physical volume of the signal, which corresponds to 123 $(\textrm{Gpc}\,h^{-1})^3$. In this case, we use the non-local Lagrangian bias expansion, with $b_1$ and $b_2$ as free parameters, which we marginalize over together with the $\{\sigma_\textrm{P},\sigma_\textrm{B},A_\textrm{P},A_\textrm{B}\}$ parameters.}%
\label{fig:Nseries_P02B02_nog2poc_5parnorm}
\end{figure}

We perform the MCMC to explore the cosmological parameter space posteriors for the same parameters and priors as for \textsc{Quijote}, (in this case the prior on $A_P$ and $A_B$ is not important, given that the shot noise signal is larger) using the \textsc{Nseries} data vectors and the covariance obtained  from the \textsc{Patchy} mocks.  The main results are  presented in Figure \ref{fig:Nseries_P02B02_nog2poc_5parnorm} and Table  \ref{tab: results}. 
In this case, the fiducial quantities for the clustering parameters are $\sigma_8=0.637,f=0.737$, resulting in $f\sigma_8=0.469$. The bias parameters $b_1,b_2$ have no value known \textit{a priori}, so they are marginalized over, together with the $\sigma_P,\sigma_B,A_\textrm{P},A_\textrm{B}$  parameters, as before.

\begin{table}[htb]
\centering
\begin{tabular}{|l||c|c|c|}
\hline

Data-vector     &  $\Delta f\sigma_8\pm1\sigma$              & $\Delta\alpha_\parallel\pm1\sigma$                   & $\Delta\alpha_\bot\pm1\sigma$         \\ \hline
\multicolumn{4}{|c|}{\textsc{Quijote} DM simulations} \\
\hline
$P_{02}$        & $-0.0044\pm0.0042$ &  $0.0027\pm0.0063$         & $-0.0023\pm0.0030$       \\ %
$P_{02}+B_0$     & $-0.0040\pm0.0038$  & $0.0047\pm0.0057$         & $-0.0017\pm0.0026$        \\ %
$P_{02}+B_{02}$ & $-0.0025\pm0.0026$  & $0.0018\pm0.0021$         & $(-4.4\pm18)\cdot10^{-4}$ \\ \hline
\multicolumn{4}{|c|}{\textsc{Nseries} galaxy simulations} \\
\hline
$P_{02}$        &  $-0.0060\pm0.0073$  & $(-8\cdot10^{-5})\pm0.00620$ & $(6.0\pm30)\cdot10^{-4}$ \\ %
$P_{02}+B_0$     & $-0.0072\pm0.0065$ &  $(3.2\pm53)\cdot10^{-4}$  & $(2.6\pm27)\cdot10^{-4}$\\ %
$P_{02}+B_{02}$ &  $-0.0077\pm0.0051$  & $0.0016\pm0.0021$         & $0.0012\pm0.0018$     \\ \hline
\end{tabular}

\caption{Performance in recovering the input cosmological parameters, for \textsc{Quijote} and \textsc{Nseries}, for the three choices of data-vector, $\{P_{02},P_{02}+B_0,P_{02}+B_{02}\}$, where $B_{02}$ denotes the combination $B_0+B_{200}+B_{020}$. The errorbars correspond to the 1$\sigma$ deviation for the effective volumes of, respectively, 100 $(\textrm{Gpc}\,h^{-1})^3$ (\textsc{Quijote}) and 80 $(\textrm{Gpc}\,h^{-1})^3$ (\textsc{Nseries}).}%
\label{tab: results}
\end{table}

We show the posterior distributions for $P_{02}+B_0$ and $P_{02}+B_{02}$, together with the constraints obtained from the  $P_{02}$ data-vector, for comparison. In the set of parameters $\{f\sigma_8,\alpha_\parallel,\alpha_\bot\}$, the addition of $B_0$ tightens the constraints by a factor of $\sim10\%$. %
This is expected, and only the inclusion of the anisotropic signal of the bispectrum (the $P_{02}+B_{02}$ data-vector) grants significant improvement in the inference of the parameters $\{\alpha_\parallel,\alpha_\bot\}$, reducing the errorbars by  $66$ and $30\%$ respectively--as in the previous case in Section \ref{sec: Quijresults}. 

There is a slight offset of $\sim 1.5\sigma$ in  $f\sigma_8$ for $P_{02}+B_{02}$.
The central value of the recovered  $f\sigma_8$ is driven by the power spectrum part of the data-vector; the added value of the bispectrum is in breaking the $f\sigma_8$ degeneracy. Given this consideration and the large volume adopted here this bias  does not invalidate the results of this paper, but this issue is further developed in Section \ref{sec: fs8 disentangle}.

In short, we find that our analysis with the \textsc{Nseries} galaxy simulations supports our claim that the bispectrum, with the modified $Z_2^\textrm{GEO}$ kernel and paired with the local Lagrangian bias expansion specified in Eq. \ref{eq: z1z2}, is a solid approach to be adopted for large volume galaxy surveys. It does not bias the results relative to the power spectrum only analysis, and increases the precision in  the inferred  the Alcock-Paczyński parameters as well as breaking the $f$--$\sigma_8$ degeneracy (see Section \ref{sec: fs8 disentangle}).

\subsection{Disentangling the \texorpdfstring{$f\sigma_8 $}{fs8} degeneracy}
\label{sec: fs8 disentangle}

It is well known that the power spectrum data-vector is sensitive to the combination $f\sigma_8$:  the two individual parameters $f$ and $\sigma_8$ are highly correlated and thus cannot be separated in a reliable way with the power spectrum alone. This is  the reason why in  Section \ref{sec: results_fs8} we do not show the $P_{02}$  results for $f$ and $\sigma_8$ separately but only show the combination $f\sigma_8$.

\begin{figure}[htb]
\centering 
\includegraphics[  width = 0.49\textwidth ]
{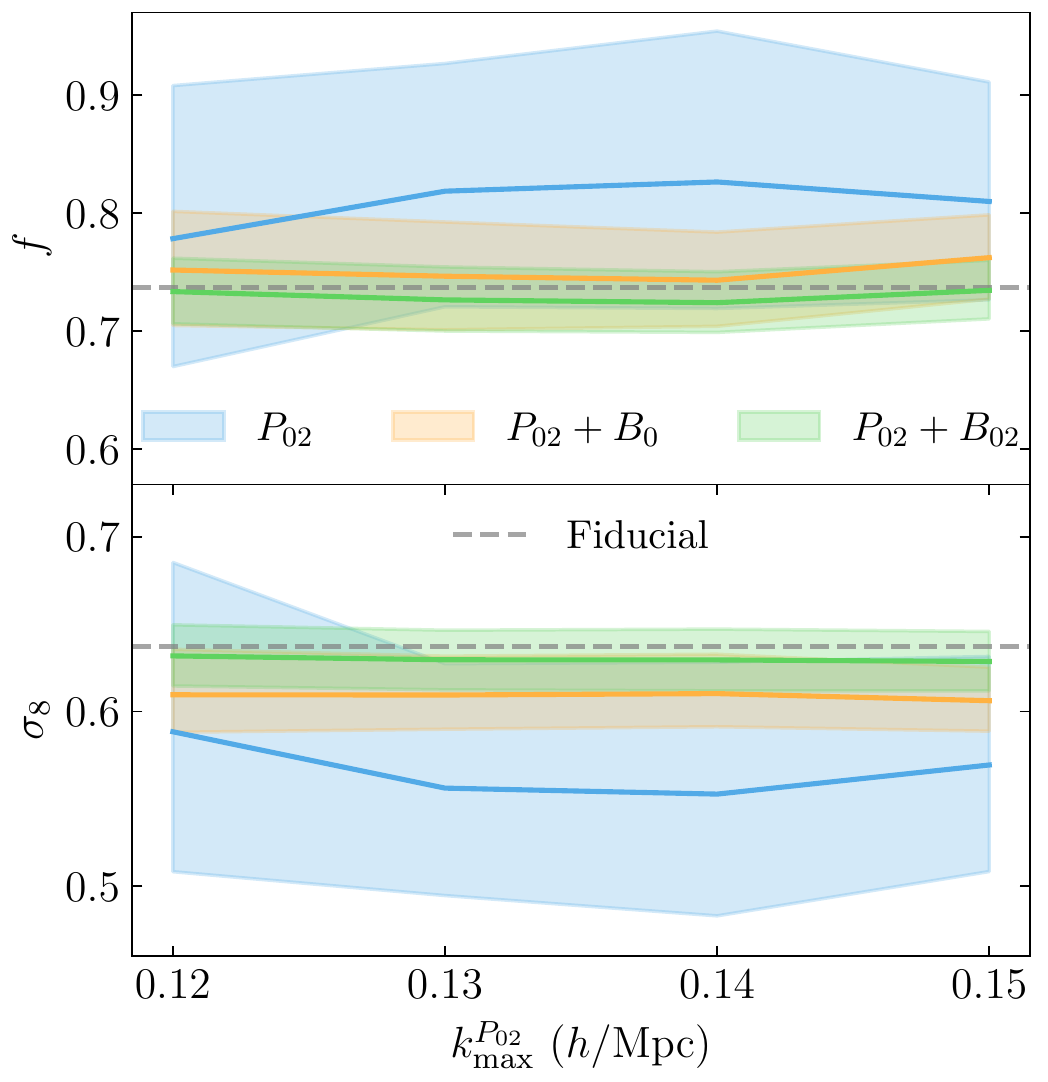}
\includegraphics[  width = 0.5\textwidth ]
{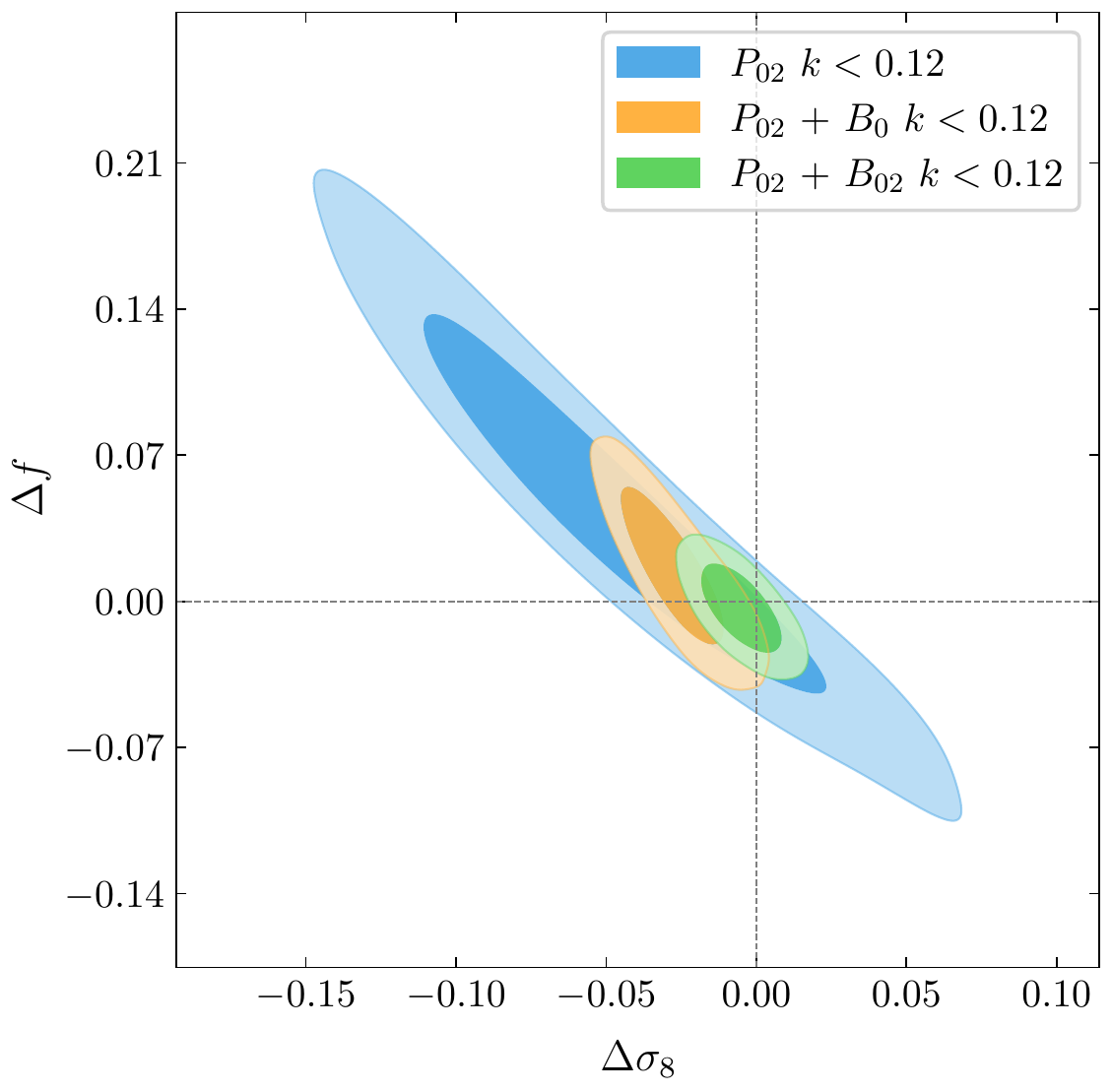}
\caption{Left panel: Recovered $f$ and $\sigma_8$ from the \textsc{Nseries} galaxy simulations, as a function of the maximum $k$-vector considered in the $P_{02}$ analysis, $k_\textrm{max}^{P}$. Right panel: Marginalized $f$ and $\sigma_8$ posteriors recovered with the \textsc{Nseries} galaxy simulations, for $k_  \textrm{max}^{P}=k_\textrm{max}^{B}=0.12$. We show the results for the data-vectors $\{P_{02},P_{02}+B_0,P_{02}+B_{02}\}$, and the errorbars correspond to an effective volume of $\sim 80$ $(\textrm{Gpc}\,h^{-1})^3$. The addition of the bispectrum monopole $B_0$ does not fully correct the bias induced by  the power spectrum, while the full data-vector $P_{02}+B_{02}$ recovers the fiducial values for $f$ and $\sigma_8$, with errorbars markedly reduced.} %
\label{fig:Nseries_P02B02_fs8post}
\end{figure}

In practice, with a large enough volume such as the one considered here, the power spectrum alone can break (weakly)  the  $f$ and $\sigma_8$ degeneracy. We find, however, that with our adopted modeling for the power spectrum multipoles (and the adopted $k_{\rm max}$),  the recovered posterior distributions underestimate  $\sigma_8$ while overestimating  $f$, as Figure \ref{fig:Nseries_P02B02_fs8post} and Table \ref{tab: fs8} show. 
The power spectrum $2\sigma$ regions for the two parameters are large -- $\Delta\sigma_8\sim 0.15,\Delta f\sim0.2$-- because the additional constraining power arises only from the non-linear corrections. The fiducial values for $f$ and $\sigma_8$ are  recovered within $2\sigma$ 
only  for $k_{\textrm{max}}\lesssim 0.12$ $h$ Mpc$^{-1}$. For higher values of the $k_\textrm{max}$ of the power spectrum, $k_\textrm{max}^{P}$, the constraints obtained with $P_{02}$ are biased.
This effect is not limited to our approach and implementation: in fact it is consistent with the findings of  \cite{maus2023comparison,maus2023inprep,lai2023inprep}.

This, along with a possible mild bias in the recovered central values of the combination $f\sigma_8$ which become appreciable only for very large volumes and in combination with the bispectrum, represents a serious limitation of the $P_{02}$ modeling  which should be corrected: in order to suitably extract the BAO feature (and thus infer the $\{\alpha_\parallel,\alpha_\bot\}$ parameters) from the power spectrum,  mildly non-linear scales, up to at least $k\sim0.15$ $h$ Mpc$^{-1}$ should be included. But state-of-the art  modelling of $P_{02}$, for this choice of $k_\textrm{max}$  and for large survey volumes, appears to introduce subtle parameters  biases, especially  on $f$ and $\sigma_8$.
We leave further investigation of this to forthcoming works, as the focus of  this paper is  the model for the bispectrum.
Nevertheless, since one of  the central motivations for including the bispectrum in the analysis is precisely to disentangle the $f-\sigma_8$ degeneracy when combined with the power spectrum,  we quantify the magnitude of the $P_{02}$-induced bias and its repercussions for the joint power spectrum-bispectrum analysis.

\begin{table}[htb]
\centering
\begin{tabular}{|l||c|c|c|c|}
\hline

  &  $k_{\rm max}^{P_{02}}=0.12$              & $k_{\rm max}^{P_{02}}=0.13$                   & $k_{\rm max}^{P_{02}}=0.14$ &$k_{\rm max}^{P_{02}}=0.15$         \\ \hline
\multicolumn{5}{|c|}{$\Delta\sigma_8\pm2\sigma$} \\
\hline
$P_{02}$        & $-0.049\pm0.088$ &  $-0.081\pm0.066$         & $-0.084\pm0.072$  &  $-0.068\pm0.062$   \\
$P_{02}+B_0$     & $-0.027\pm0.024$  & $-0.028\pm0.021$         & $-0.027\pm0.020$  &   $-0.031\pm0.018$   \\
$P_{02}+B_{02}$ & $-0.005\pm0.018$  & $-0.007\pm0.017$         & $-0.007\pm0.018$& $-0.008\pm0.017$\\
\hline
\multicolumn{5}{|c|}{$\Delta f\pm2\sigma$} \\
\hline
$P_{02}$        &  $0.04\pm0.12$  & $0.08\pm0.10$ & $0.09\pm0.11$&$0.073\pm0.092$ \\ 
$P_{02}+B_0$     & $0.014\pm0.048$ &  $0.010\pm0.046$  & $0.006\pm0.040$& $0.025\pm0.036$  \\ 
$P_{02}+B_{02}$ &  $-0.004\pm0.028$  & $-0.011\pm0.027$         & $-0.013\pm0.026$ &  $-0.003\pm0.025$  \\ \hline
\end{tabular}

\caption{Recovered constraints for the parameters $\sigma_8$ (upper panel) and $f$ (lower panel) for the \textsc{Nseries} simulations, as a function of $k_{\rm max}^{P}$, while $k_{\rm max}^{B}$ is fixed to 0.12.  We display the three choices of data-vector $\{P_{02},P_{02}+B_0,P_{02}+B_{02}\}$, where, as stated throughout the paper, $P_{02}$ and $B_{02}$ denote the combinations $P_0+P_2$ and $B_0+B_{200}+B_{020}$ respectively. The errorbars correspond to the 2$\sigma$ deviation for the \textsc{Nseries} effective volume, 80 $(\textrm{Gpc}\,h^{-1})^3$. }
\label{tab: fs8}
\end{table}

As Figure \ref{fig:Nseries_P02B02_fs8post} and Table \ref{tab: fs8} show, when
adding the bispectrum monopole to the data-vector, %
the peaks of the posterior distributions  approach the true values. When the bispectrum quadrupoles are  also included, in the $P_{02}+B_{02}$ case, the offset induced by  the power spectrum is fully corrected.
This offers a further motivation for  utilizing the full $B_{02}=B_0+B_{200}+B_{020}$ data-vector.

\subsubsection{Additional considerations and robustness tests}

A possible source of error in the power spectrum modeling could be the integration of the loop terms. In theory the range of the integrals in the 1 and 2-loop terms is from $k=0$ to infinity, while in practice we use a reasonable upper cutoff scale of $k=1$ $h$ Mpc$^{-1}$. We investigate whether increasing the cutoff scale affects  the recovered parameters in Appendix \ref{App: Nseries}. There, we show that this effect is absorbed by the nuisance parameter $\sigma_\textrm{P}$, so we confirm that for this application, this approximation is adequate and does not introduce bias in the analysis.  

Moreover, while we fit the bispectrum kernel to the data vector of bispectrum monopole and quadrupoles, parameters biases could be introduced  if redshift space distortions induced  a preference for different kernel parameters of Equation \ref{eq: Z2_geo} between the bispectrum monopole and quadrupoles.
We explore this extension, which we refer to as `2-kernels' modification, in   Appendix \ref{App: Nseries} where 
we conclude that this preference is not significant; hence we do not consider further any of the `2-kernels' modifications, which we consider would be over-fitting.

\subsection{Testing strong growth factor variations}
\label{sec: CW}
We use the CW simulations to explore the performance of  the GEO-FPT model with $\Lambda$CDM cosmologies featuring significantly different values of $\Omega_m$ from the ones where the kernels have been calibrated. This effectively tests the model's performance for significantly different growth factors. 

The single CW simulations with  matter densities  $\Omega_m= \{ 0.2, 0.27, 0.4, 1.0\}$ feature the same initial conditions\footnote{For the fiducial cosmology with  $\Omega_m=0.27$ there are 160 simulations in total, of which one has the same initial conditions as for the other cosmologies.}.
As  further discussed in  Appendix \ref{app: Quijote}, the CW simulations are only sufficiently accurate to study relative effects (variations with respect to the fiducial, $\Omega_m=0.27$ simulation) and only up to $k_{\rm max}=0.11$.

\begin{figure}[htb]
\centering 
\includegraphics[ width = \textwidth ]
{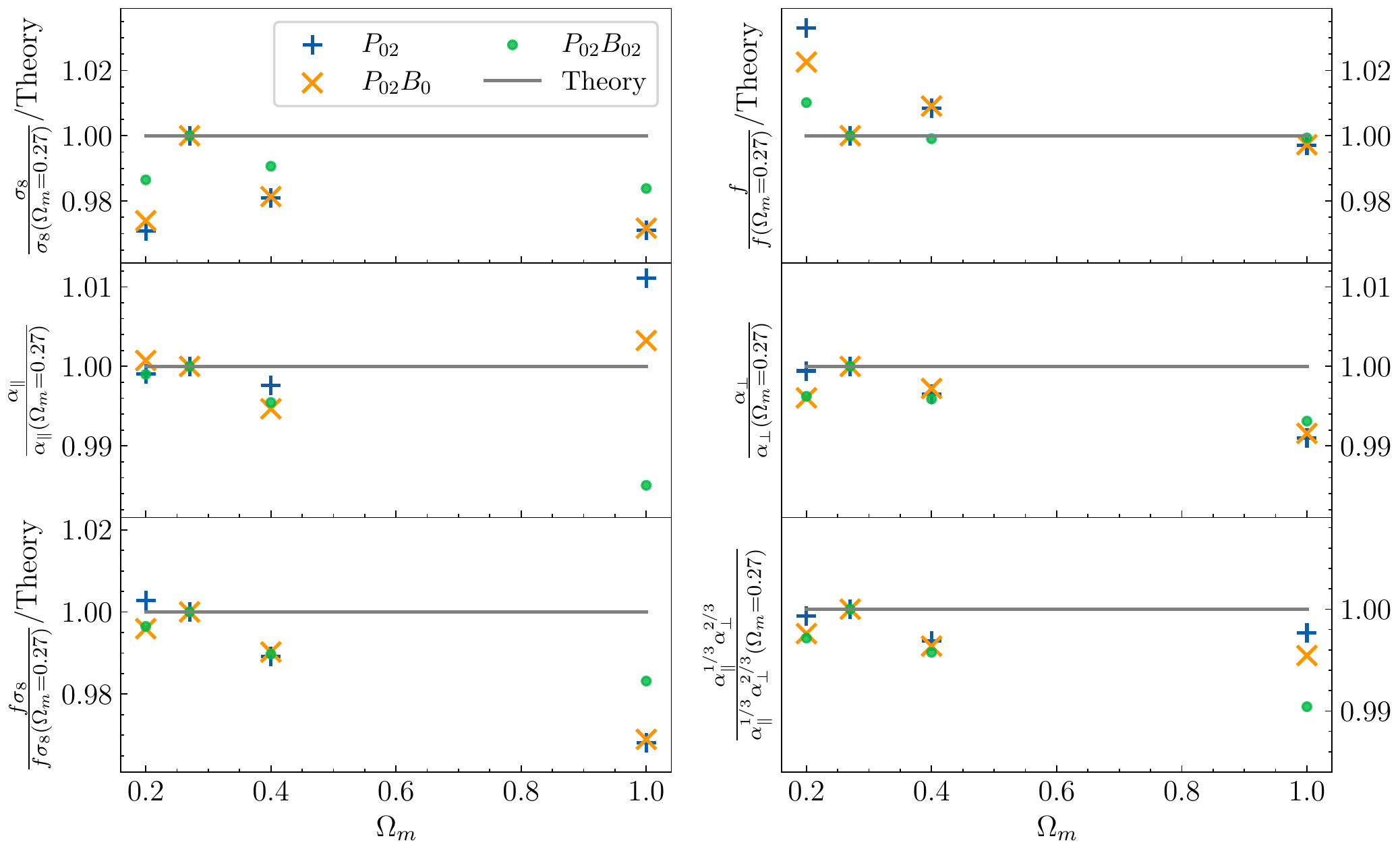}
\caption{Increase of systematic errors as a function of distance from the fiducial $\Omega_m=0.27$ case --where the cosmology is closest to the \textsc{Quijote} simulations, in which the bispectrum effective kernel $Z_2^{\textrm{GEO}}$ has been calibrated. The parameters are inferred from the dark matter CW simulations for $\Omega_m=0.2,0.27,0.4,1$, one box of $(2.4 \textrm{Gpc}\,h^{-1})^3$ each. The recovered parameters ratios (to the fiducial  cosmology simulation)  are normalized  by the theoretical value in the case of $f$, $\sigma_8$ and $f\sigma_8$ for clarity. As seen before, when $f$ and $\sigma_8$ are treated as independent parameters,  the bias in the recovered values is driven by the power spectrum.}
\label{fig:CWrun1_Omall_rpt_06711}
\end{figure}

We perform cosmological parameters inference with the \textsc{Quijote}-calibrated GEO-FPT bispectrum, for  each of the single simulations that share the same initial conditions, using the  more conservative value of $k_\textrm{max}=0.11$. In all cases, we adopt the covariance estimated from the fiducial set of \textsc{Quijote} simulations; this assumes, as usually done, that the covariance does not  change  radically with cosmology.

The maximum likelihood parameters, relative to those of the fiducial simulation, are shown in Figure \ref{fig:CWrun1_Omall_rpt_06711}. There is an increase of systematic errors with distance from the fiducial case (where the cosmology is closest to the  \textsc{Quijote} simulations). 
The systematic shifts in the various combinations of the AP parameters are driven by the power spectrum, but are  below  0.3\% except,  not unexpectedly, for $\Omega_m=1$, where they grow to 1\%. 
The deviations on the $f\sigma_8$ parameter are within 1\% in the cases with $\Omega_m=0.2-0.4$ for all data-vectors,  and are also driven by the power spectrum. As seen in Section \ref{sec: fs8 disentangle},  the full data-vector $P_{02}+B_{02}$ helps  breaking the $f-\sigma_8$ degeneracy correctly, reducing the bias driven by $P_{02}$. This bias remains below 2\%  even at  $\Omega_m=1$.

While not unexpected, given that the SPT bispectrum kernels have a very weak cosmology dependence,  these tests  demonstrate that  GEO-FPT bispectrum (calibrated on a fiducial cosmology) works well  even for cosmologies that deviate %
from the Planck-preferred one. The small residual biases (which only become visible for such large volumes) appear to be driven by limitations in the modeling of the power spectrum.

\section{Conclusions}
\label{sec: conclusions}
Higher-order statistics such as the bispectrum are promising tools to extract cosmological information beyond that encoded in the power spectrum. Having an accurate  enough bispectrum model is one of the main challenges  that an analysis of the bispectrum of a galaxy redshift survey  involves.

We have presented the GEO-FPT model for the redshift space bispectrum monopole and quadrupoles. GEO-FPT is a phenomenological modification of the tree-level SPT bispectrum consisting in a modulation of the $Z_2$ SPT kernel by a function with  5 free parameters that we have calibrated on the \textsc{Quijote} dark matter simulations for a k-range of  $0.02<k[h\,\textrm{Mpc}^{-1}$]<0.12,  and redshift range $z=0.5-2.0$.
This $k$-range provides access to the information contained in the mildly non-linear regime. 
In the $k$-range of interest, GEO-FPT reproduces the bispectrum monopole from N-body simulations at $\lesssim3\%$ level, and the quadrupoles at $\lesssim 30\%$,  well within the 1$\sigma$ statistical errors for a survey  volume of 100 $(\textrm{Gpc}\,h^{-1})^3$.

The  GEO-FPT performance in recovering  the cosmological parameters $\{\sigma_8,f,\alpha_\parallel,\alpha_\bot\}$, is validated in three different sets of synthetic catalogs: the same \textsc{Quijote} simulations where it had been calibrated; the \textsc{Nseries} galaxy simulations, featuring a WMAP-compatible cosmology; and the CW dark matter simulations, with values of $\Omega_m=0.2,0.27,0.4,1,$ at fixed $\Omega_mh^2.$
GEO-FPT provides a markedly improved performance (in terms of accuracy and $k$-range) compared to other state-of-the art bispectrum models.

Both the \textsc{Quijote} and \textsc{Nseries} analyses are carried out for effective volumes of 100 and 80 $(\textrm{Gpc}\,h^{-1})^3$ respectively, far exceeding the target of upcoming surveys. This is a stringent test for the systematic errors of the GEO-FPT model, which recovers  the cosmological  parameters within the narrow statistical error correspondent to the adopted cosmological volume. In particular, the \textsc{Nseries} results show that the GEO-FPT model is readily applicable to biased tracers, such as galaxies and quasars. The test on the  CW simulations (which feature strongly different growth factors and $\Omega_m$) %
 further confirms that the GEO-FPT model remains valid for a wide range of cosmologies.

Including  the bispectrum monopole and quadrupoles  in a joint analysis with the power spectrum monopole and quadrupole not only improves dramatically the  errorbars on $\alpha_\parallel$ and $\alpha_\bot$ (reducing them by respectively 66 and 30\%), but also  breaks  the $f-\sigma_8$  quasi-degeneracy present in the power spectrum.
The state-of-the-art modeling of the anisotropic galaxy power spectrum in redshift space used here is shown to be the main limitation in recovering unbiased estimates for the individual  $f$ and $\sigma_8$ parameters. This should be an important objective of future work.
In view of an application to forthcoming galaxy surveys, further work is also needed to develop a suitable  modeling on the survey  window function on   the galaxy bispectrum monopole and quadrupoles (see e.g. \cite{sefusatti2022bispectrum,porciani2023window} for pioneering efforts in this direction).

The calibrated model and related data vectors and covariances are  made publicly available at \url{https://github.com/serginovell/Geo-FPT}. Two options are provided: a calibration in view of the use of the full bispectrum monopole and quadrupoles data vector and one that includes the bispectrum monopole only, for cases where the inclusion of the multipoles is unfeasible.

\section*{Acknowledgements}
SNM acknowledges funding from the official doctoral program of the University of Barcelona for the development of a research project under the PREDOCS-UB grant.
HGM acknowledges support through the program Ram\'on y Cajal (RYC-2021-034104) of the Spanish Ministry of Science and Innovation. 
LV, DG and HGM acknowledge support of European Union’s Horizon 2020 research and innovation programme ERC (BePreSySe, grant agreement 725327).

Funding for this work was partially provided by the Spanish MINECO under project PGC2018-098866-B-I00MCIN/AEI/10.13039/501100011033 y FEDER ``Una manera de hacer Europa", and the ``Center of Excellence Maria de Maeztu 2020-2023'' award to the ICCUB (CEX2019-000918-M funded by MCIN/AEI/10.13039/501100011033).

This work has made extensive use of the following publicly available codes: \href{https://lesgourg.github.io/class_public/class.html}{\textsc{Class}}, \href{https://emcee.readthedocs.io/en/stable/index.html}{\textsc{Emcee}}, \href{https://github.com/hectorgil/PTcool}{\textsc{PTcool}}, \href{https://www.gnu.org/software/gsl/}{GSL}, \href{https://github.com/hectorgil/Rustico}{\textsc{Rustico}},  \href{https://scipy.org/}{\textsc{SciPy}}, \href{https://numpy.org/}{\textsc{NumPy}}, \href{https://getdist.readthedocs.io/en/latest/}{\textsc{GetDist}}, \href{https://matplotlib.org}{\textsc{Matplotlib}}, \href{https://fftw.org/}{FFTW}. We are grateful  to the developers who made these codes public.

\appendix
\section{Covariance matrix estimation and results}
\label{app: cov}
Throughout this work, all the covariance matrices have been estimated from simulations in the traditional way as follows. Let $\textbf{D}$ be the data-vector of interest, $\textbf{D}_i$ be the measured data-vector in the $i$-th simulation, and $n$ the number of simulations. The covariance matrix $\textbf{C}$ is obtained as
\begin{equation}
    \textbf{C}=\frac{1}{n-1}\sum_{i=1}^n(\textbf{D}_i-\overline{\textbf{D}})(\textbf{D}_i-\overline{\textbf{D}})^T,
\end{equation}
where $\overline{\textbf{D}}$ is the mean across all realizations,
\begin{equation}
    \overline{\textbf{D}}=\frac{1}{n}\sum_{i=1}^n \textbf{D}_i.
\end{equation}

\begin{figure}[htb]
\centering 
\includegraphics[ width = 0.5\textwidth ]
{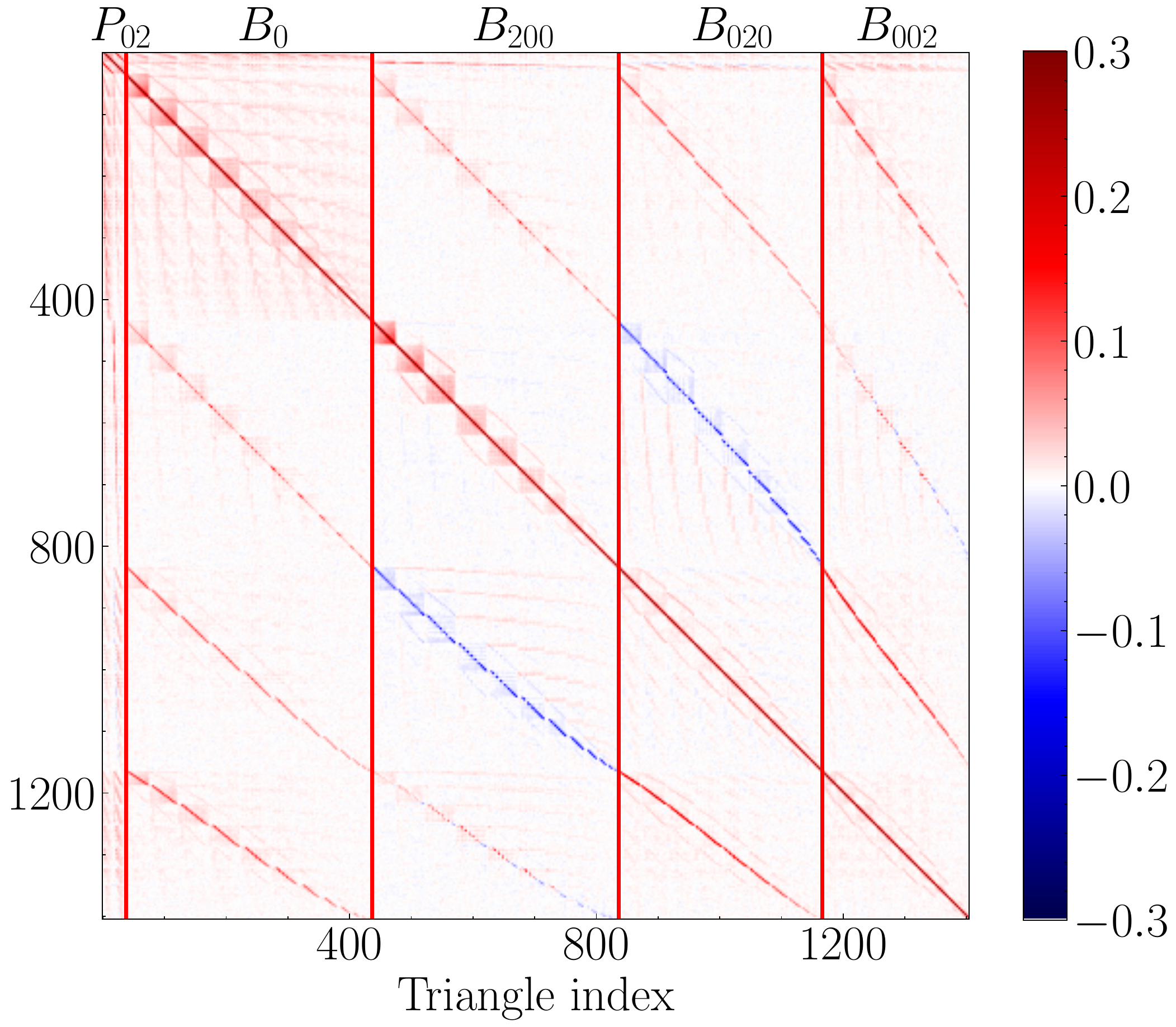}
\includegraphics[trim=0 -0.86cm 0 0,width = 0.47\textwidth ]
{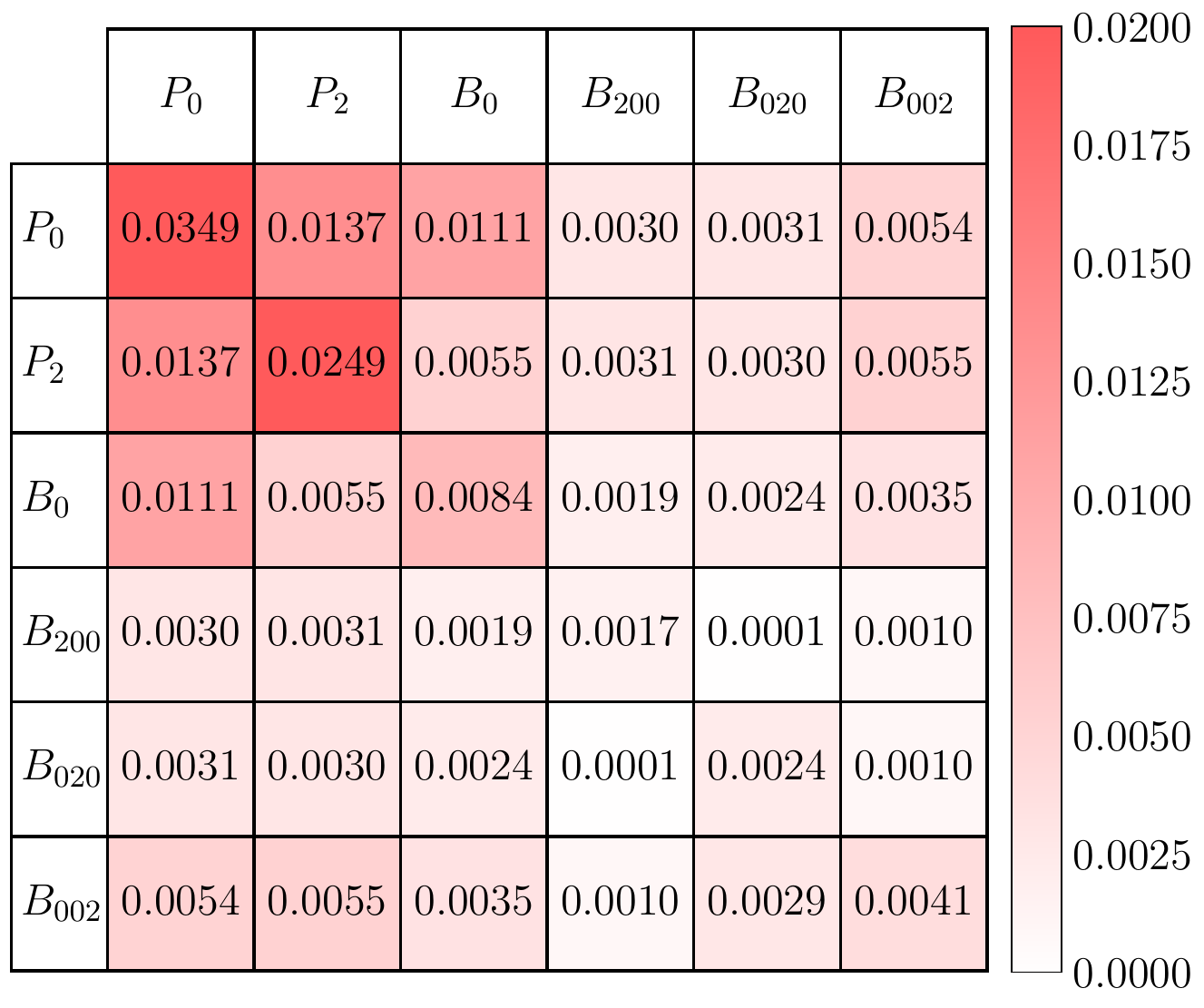}

\caption{Left panel:  reduced covariance matrix of the data-vector formed by $\{P_0,P_2,B_0,B_{200},B_{020},B_{002}\}$ at $z=0.5$: extended version of Figure \ref{fig:B02_Quijfid_red_fit100_sm}. Together with the structure within the bispectrum data-vector, we can observe  how the power spectrum is clearly correlated with the bispectrum, especially with the bispectrum monopole $B_0$. Right panel: median of all correlation terms between each pair of data-vectors. At each cell, we show the median value of all the elements in the box of the reduced covariance matrix that relate the data-vectors in the corresponding row and column. }
\label{fig: fullcov}
\end{figure}

The reduced covariance matrix of the bispectrum data-vector used in this work is  plotted in Figure \ref{fig:B02_Quijfid_red_fit100_sm}. For  completeness, we show in Figure \ref{fig: fullcov} the covariance matrix for the full data-vector used in the cosmological parameter inference plus the $B_{002}$ multipole, which includes the auto and cross terms involving the power spectrum monopole and quadrupoles. The left panel illustrates the fine details of the full structure of the covariance matrix, while the right panel quantifies the auto and cross-correlations in blocks by showing the median value of all coefficients in the corresponding box. The values correspond to the reduced covariance matrix, so that the terms concerning the power spectrum and bispectrum have comparable value. Of course, this is an extreme compression of all the information present in the covariance matrix, but it  qualitatively illustrates the importance of the cross-correlations compared with the corresponding auto-correlations. 

The errorbars that result from using a covariance estimated with $n$ simulations  correspond to an equivalent volume of one such simulation, $V_\textrm{1sim}$. When doing parameter inference, if we use the mean of $n$ simulations as the measured data-vector, this signal corresponds to that of the cumulative volume covered by the $n$ simulations.  In  fitting  the free parameters of our formula, $\{f_1,...,f_5\}$, we  rescale the estimated covariance matrix by a factor of $n$, thus effectively transforming the errorbars from corresponding to a volume of $V_\textrm{1sim}$ to $n\times V_\textrm{1sim}$.

In particular, when fitting $\{f_1,...,f_5\}$ we rescale the covariance estimated with the \textsc{Quijote} simulations by the number of realizations at each redshift, 8000 at $z=0.5$ and 4000 at $z=1,2.$ These errorbars can be seen in Figures \ref{fig:Z2kern5par_redCW_mpc100}, \ref{fig:matpk_QvCW_comp2LRPT_facc}. 

It is only in the cosmological parameter recovery in the  \textsc{Quijote} sims, in Figures \ref{fig:Quijote_P02B02_nog2poc_5parnorm} and \ref{fig:Quijote_P02B02_fs8post} (right panel), that we opt for rescaling the covariance by a factor different than the number of simulations. In particular we set   $n=100$ in this case,  corresponding to a volume of $100 \textrm{ (Gpc}\,h^{-1})^3.$

Similarly, in doing the cosmological parameter exploration for  \textsc{Nseries} (Figures \ref{fig:Nseries_P02B02_nog2poc_5parnorm}, \ref{fig:Nseries_P02B02_nog2poc_5parnorm12kern}), we rescale the covariance by a factor of 7, to  obtain errorbars for  an equivalent physical volume of $7\times (2.6\textrm{ Gpc}\,h^{-1})^3=123\textrm{ (Gpc}\,h^{-1})^3$ (effective volume of $80  \textrm{ (Gpc}\,h^{-1})^3$). We compute the effective volume from a given physical volume as in \cite{tegmark1997measuring}. This is in line of the main goal of this paper, which is to show that our proposed model is appropriate for cosmological volumes $\leq 100\textrm{ (Gpc}\,h^{-1})^3.$

We use the covariance obtained with \textsc{Quijote} also for the analysis performed on the CW simulations, assuming that the covariance does not strongly depend on cosmology. To match the volume of each CW realization, we rescale the covariance by $n=\left(L_\textrm{box}^\textrm{CW}/L_\textrm{box}^\textrm{Quijote}\right)^3$. This is however of negligible importance for the role of the CW simulations in this paper, since we only consider the ratios between maximum likelihood parameters, as seen in Section~\ref{sec: CW}.

\section{\textsc{Quijote} complementary results}
\label{app: Quijote}
\begin{figure}[htb]
\centering 
\includegraphics[ width = 0.9\textwidth ]
{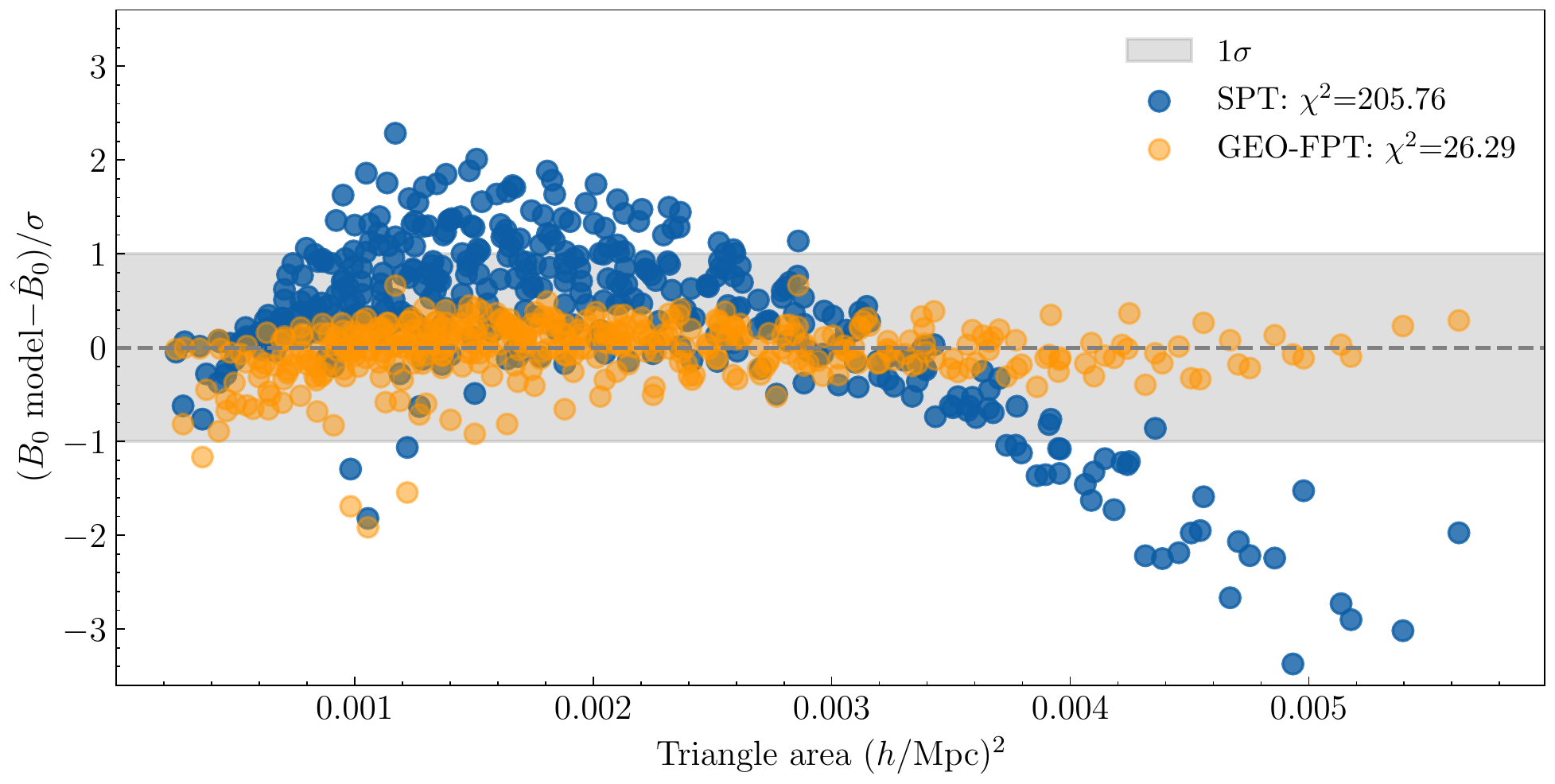}
\includegraphics[ width = 0.9\textwidth ]
{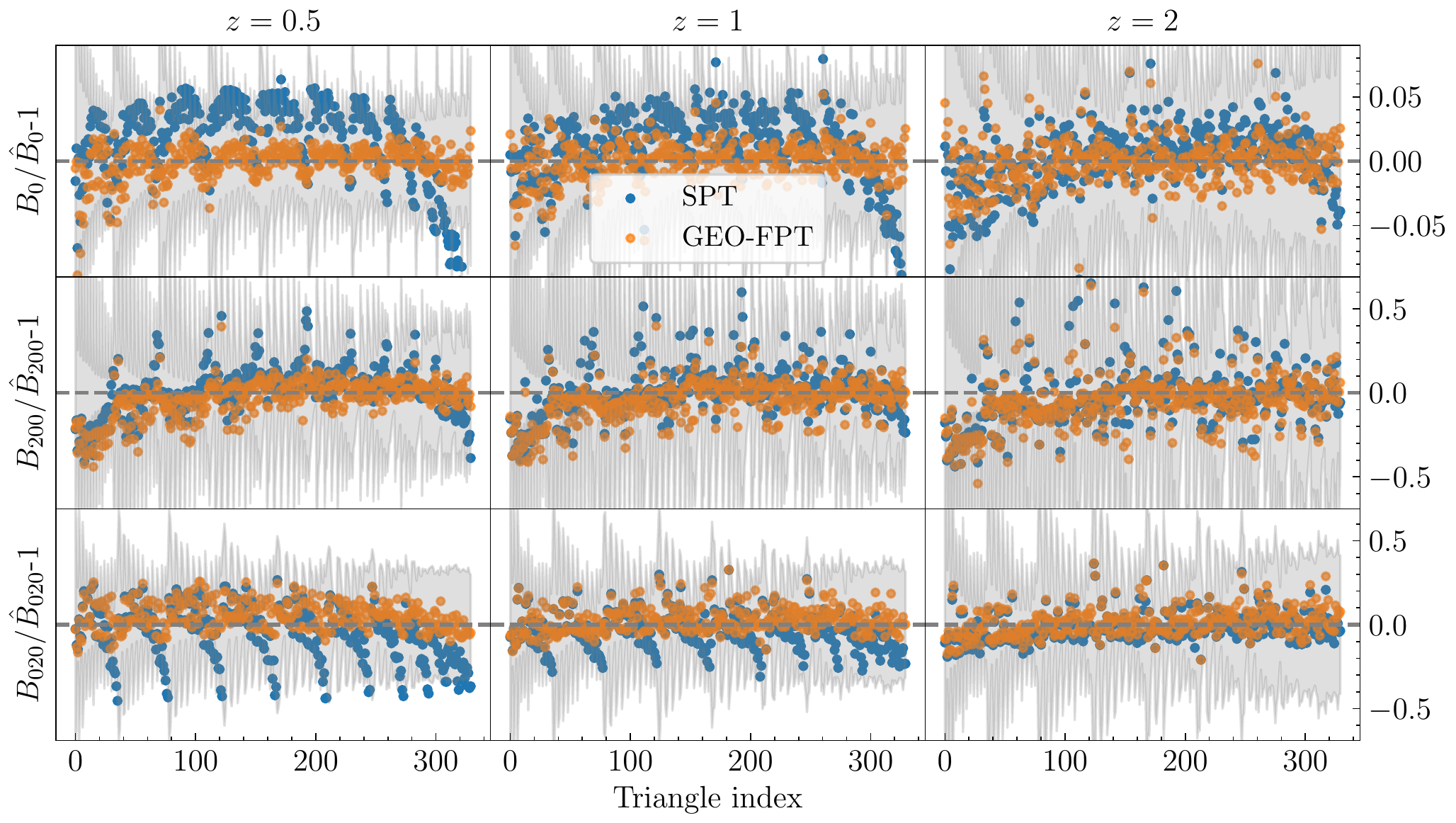}
\caption{Performance of GEO-FPT (orange symbols) and tree-level SPT (blue symbols) in describing the measured bispectrum monopole signal in the dark matter \textsc{Quijote} simulations. The upper panel is similar to that shown in  the lower panel of Figure \ref{fig:B0_Quijfid_red_spt}, where the difference between bispectrum model and measurements is shown as a function of the triangle area (each dot in the figure corresponds to a bispectrum triangle). In this case we only display the $z=0.5$ outputs for clarity. The total number of configurations is  399 for triangles with $k_i\leq0.12\,h\,{\rm Mpc}^{-1}$, binned such that $\Delta k_i=1.1\,k_{\rm f}$. For the tree-level SPT model only the FoG damping term is freely varied, and consequently the  number of degrees of freedom is $399-1$; whereas for the GEO-FPT 5 additional $f_i$ parameters (see Table~\ref{tab: fit_quij}) have been calibrated to these simulations, the number of degrees of freedom is $399-6$. The reported $\chi^2$ values in the legend correspond to a covariance with an associated effective volume of $100\,{\rm Gpc}\,h^{-1}$, whereas the signal has an associated volume of $8000\,{\rm Gpc}\,h^{-1}$, hence the `low' $\chi^2$ in both cases. The lower panels are analogous to the right panels in Figure~\ref{fig:B02_Quijfid_red_fit100_sm}, where the comparison is for the multipoles $B_0,B_{200},B_{020}$ for the redshifts $z=0.5,1,2$. The GEO-FPT model shows a much better agreement with the simulated data even  considering the additional adjustable parameters of the model.}
\label{fig:B0_Quijfid_red_sptfpt}
\end{figure}

In Figure~\ref{fig:B0_Quijfid_red_sptfpt} a comparison between the model presented in this paper (GEO-FPT, Eq.~\ref{eq:bispredshiftsp}), and the standard tree-level Standard Perturbation Theory (SPT, Eq.~\ref{eq:bispredshiftspGEO}) is shown. Such comparison is presented  as a function of the triangle area for $z=0.5$ for the bispectrum monopole (top panel), similarly to what was already shown in Figure~\ref{fig:B0_Quijfid_red_spt} of Section~\ref{sec: theory}, as a function of the triangle index for $z=0.5,\,1,\,2$, and for the bispectrum monopole and quadrupoles, as indicated (bottom panels). We report a substantial improvement of the performance of the GEO-FPT model over SPT model for the explored triangle configurations ($k_i\leq0.12\,h\,{\rm Mpc}^{-1}$). This improvement is increasingly  evident  with  increasing  area of the triangle (in Fourier space). This is the expected trend; when the area of the $k$-triangle increases, also the $k$-sides increase, corresponding to  smaller scales, where the tree-level SPT model is expected to fail more dramatically. These improvements can be quantified by the $\chi^2$ values, as reported in the legend. As in Figure \ref{fig:B0_Quijfid_red_spt}, the grey band displays the $1\sigma$ error corresponding to a volume of $100\,{\rm Gpc}\,h^{-1}$, where the $\hat{B}_0$ signal comes from a total volume of $8000\,({\rm Gpc}\,h^{-1})^3$. 

In the following subsections we address the power suppression effect between theory and simulations (Appendix~\ref{app: Quijote power sup}) and quantify the effect of $P_{02}$ in the recovered constraints with \textsc{Quijote} for $z=1,\,2$ when the $f-\sigma_8$ degeneracy is broken (Appendix~\ref{app: Quijote fs8}).

\subsection{Comparison Quijote-CW simulations}
\label{app: Quijote power sup}
\begin{figure}[htb]
\centering 
\includegraphics[ width = 0.47\textwidth ]
{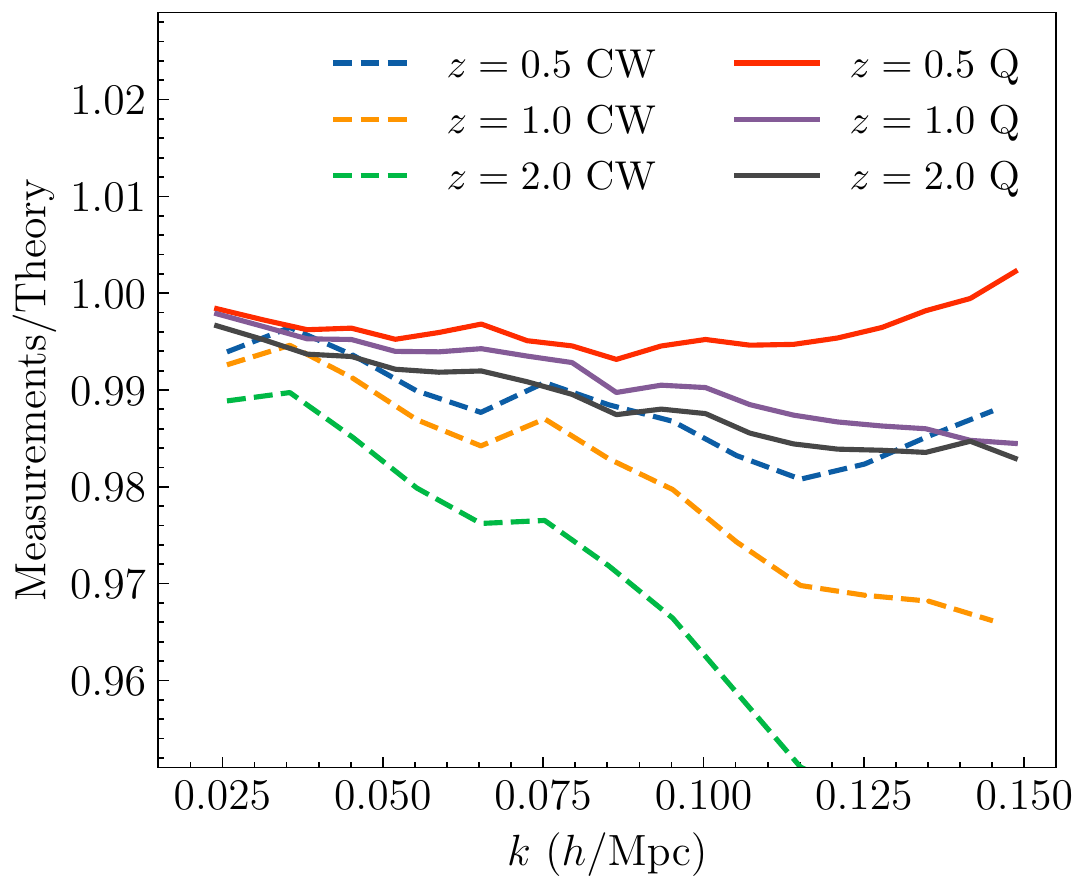}
\includegraphics[  width = 0.52\textwidth ]
{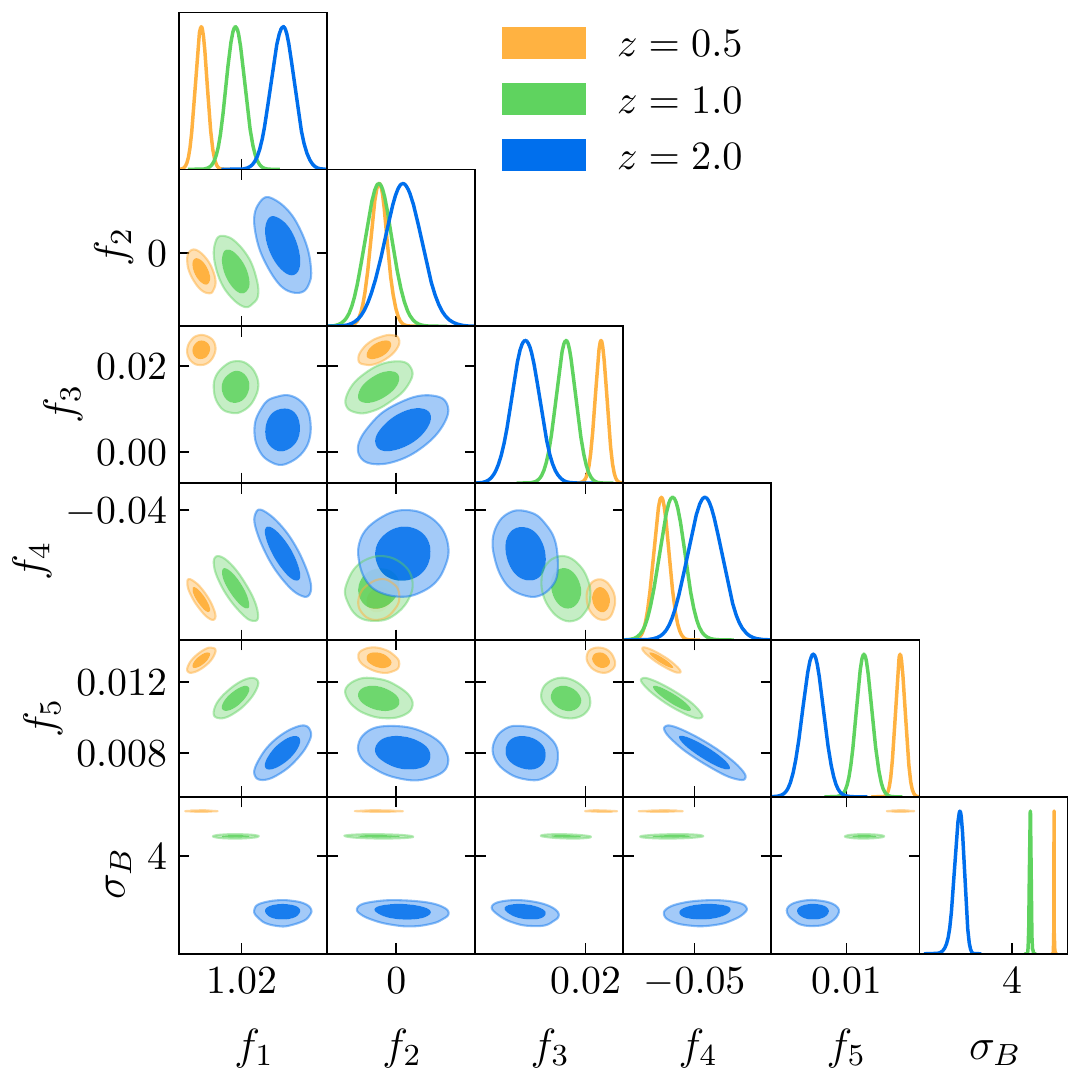}
\caption{Left panel: Loss of power of both \textsc{Quijote} and CW simulations, at $z=0.5,1,2$. The ratios between theoretical and measured non-linear matter power spectrum are shown. The theoretical model is 2L-RPT, as in the main analysis, computed for the simulation cosmology. While in the case of \textsc{Quijote} the loss of power is almost always $<1\%$, in the case of CW  the effect is much larger especially at z=1 and 2 and at $k\gtrsim 0.1$ at $z=0.5$. Right panel: Posterior distributions for the calibration of the $Z_2^{\textrm{GEO}}$ kernels at the three redshifts of interest, $z=0.5,1,2$, where the covariance has been rescaled to match the full volume of the signal (8000 $(\textrm{Gpc}\,h^{-1})^3$ at $z=0.5$ and 4000 $(\textrm{Gpc}\,h^{-1})^3$ at $z=1,2$). This is  complementary to Figure \ref{fig:Z2kern5par_redCW_mpc100}, since this Figure shows  the parameters' correlations. The strongest correlation is between $f_1,f_4,f_5$, while $f_3$ appears to be very orthogonal to the other parameters, as well as the nuisance parameter $\sigma_B$ (as expected).  }
\label{fig:matpk_QvCW_comp2LRPT_facc}
\end{figure}

It is well known that N-body simulations feature a loss of power at small scales, dependent on the initial conditions redshift $z_{\textrm{ini}}$ and the time-stepping/mass resolution \cite{scoccimarro1998transients,crocce2006transients,michaux2021accurate}. In the left panel of Figure \ref{fig:matpk_QvCW_comp2LRPT_facc} we quantify how this affects  the non-linear matter power spectrum, which is indicative as well of its effect on higher-order statistics such as the bispectrum. The Figure shows the ratio $P_{\textrm{measured}}/P_{\textrm{theory}}$, where $P_{\textrm{measured}}$ is the estimated quantity from the real space simulations and $P_{\textrm{theory}}$ is computed for the fiducial model following Equation \ref{eq: Pnl_matter}.

This power suppression  in the power spectrum,  which is stronger at higher redshifts and for the CW simulations, is visible, possibly due to CW having a lower initial conditions redshift and different resolution/time-stepping settings. While for the \textsc{Quijote} simulations the loss of power is contained within 1\% for the  $k$-range used in the bispectrum analysis, the CW measurement of the matter power spectrum at $z=0.5$ is already deviating 1$\%$ from the theory at $k=0.08$ $h$ Mpc$^{-1}$. This worsens with  $k$ and redshift.

Systematic errors of this magnitude have been acceptable until the recent past, as they were below the statistical errors. However, in this work,  given the large volume considered ( $\sim100$ $(\textrm{Gpc}\,h^{-1})^3$), the effects of this power suppression may become relevant. 
The power suppression of the CW simulations precludes us from repeating a similar analysis to that of Sections \ref{sec: Quijresults} and \ref{sec: Nseries}. Therefore, instead of evaluating the recovered cosmological parameter posteriors of every case separately, in Section \ref{sec: CW} we  show the results at each cosmology relative to those for the fiducial one, while setting a more conservative $k$-range of $0.02<k[h\,\textrm{Mpc}^{-1}]<0.11$ for the bispectrum.

\subsection{Further details on the posterior distributions for \textsc{Quijote} simulations}
\label{app: Quijote fs8}
The right panel of Figure \ref{fig:matpk_QvCW_comp2LRPT_facc} shows the posterior distributions for the kernel parameters $\{f_1,...,f_5\}$, obtained as described in Section \ref{sec: methodology}. The degeneracies between each pair of parameters are now visible, with the $f_1-f_4-f_5$ degeneracy being the most apparent.

\begin{figure}[htb]
\centering 

\includegraphics[ width = 0.49\textwidth]
{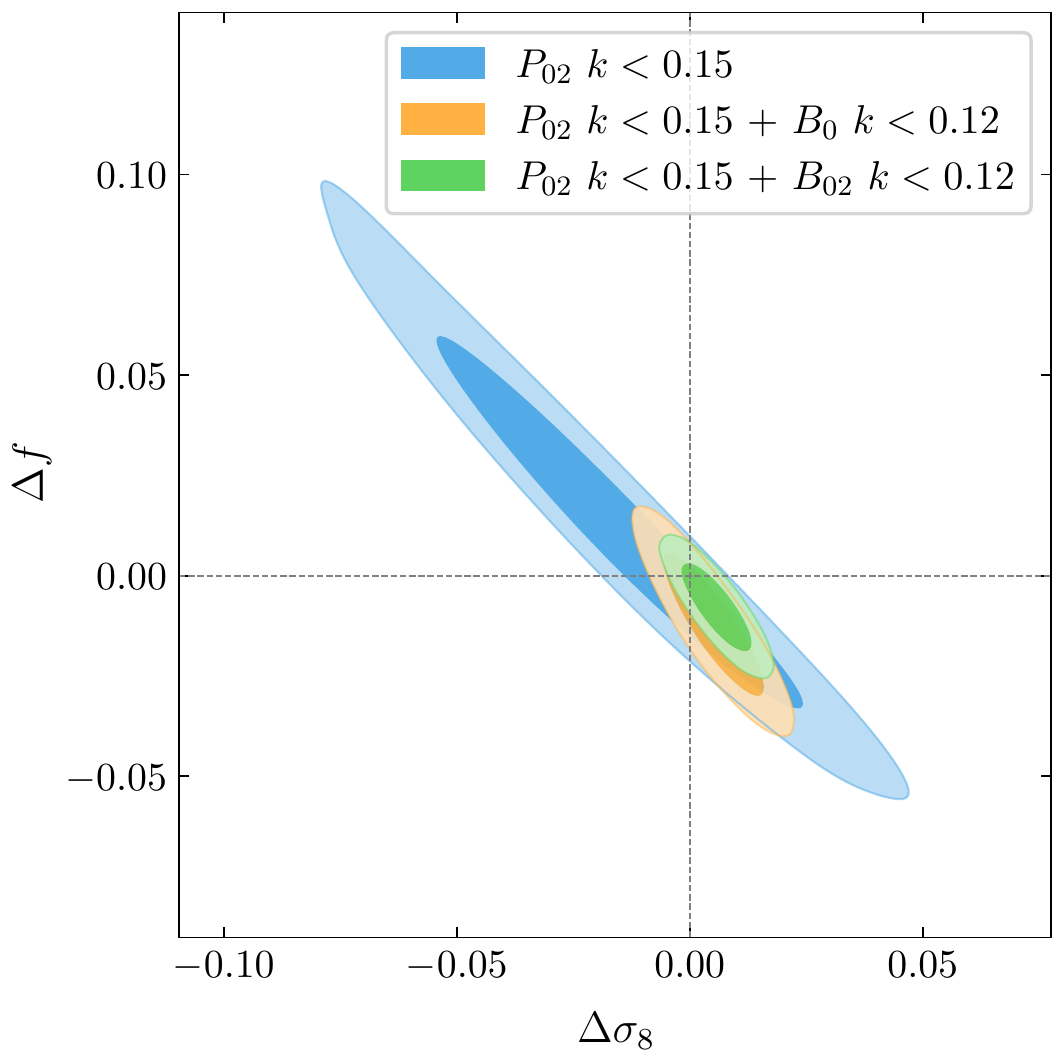}
\includegraphics[  width = 0.49\textwidth ]
{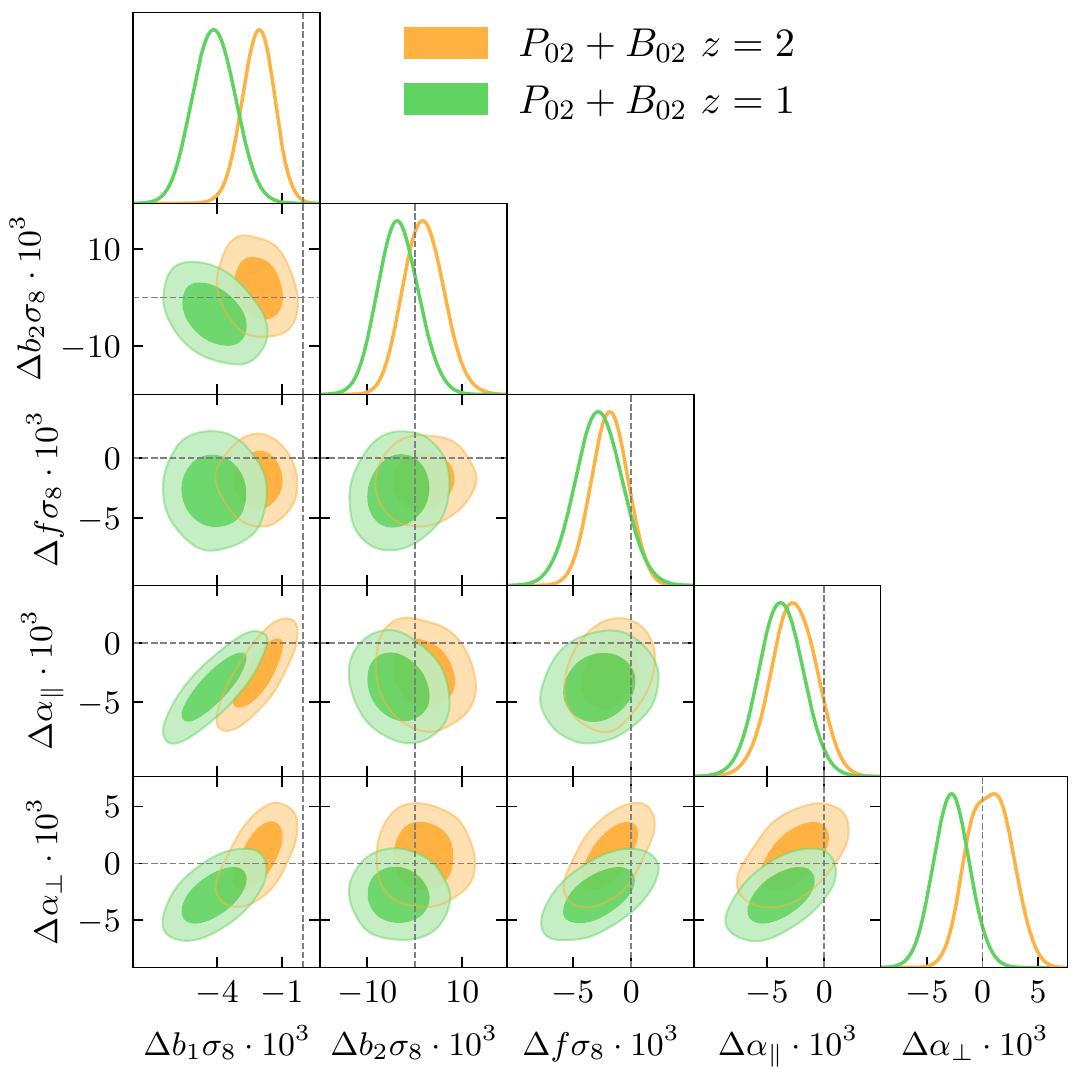}
\caption{Left panel: 2D posterior distribution for the pair of parameters $f$ and $\sigma_8$ for the \textsc{Quijote} simulations  at $z=0.5$. Differently than in Figure \ref{fig:Nseries_P02B02_fs8post} (\textsc{Nseries} galaxy mocks), the result for \textsc{Quijote} simulations is unbiased for both the power spectrum and the combined power spectrum-bispectrum data-vectors. Right panel: Main cosmological parameters recovered from the \textsc{Quijote} fiducial simulations, at redshifts $z=1,2$, with the variables multiplied by $10^3$. All settings are as in Figure \ref{fig:Quijote_P02B02_nog2poc_5parnorm}, except that here for the two redshift snapshots we show results for the full data vector.} 
\label{fig:Quijote_P02B02_fs8post}
\end{figure}

In the left panel of Figure~\ref{fig:Quijote_P02B02_fs8post}, we show the 2D constraints for $f$ and $\sigma_8$ at $z=0.5$ for the data-vectors $\{P_{02},P_{02}+B_0,P_{02}+B_{02}\}$. As in  Section \ref{sec: Quijresults}, the adopted maximum k-vectors are $k_\textrm{max}^{P}=0.15$ for the power spectrum and $k_\textrm{max}^{B}=0.12$ for the bispectrum. For a distribution of dark matter tracers, all constraints are unbiased at $\sim1\sigma$ level for a volume of 100 $(\textrm{Gpc}\,h^{-1})^ 3$. Comparison with Section \ref{sec: fs8 disentangle} indicates that the bias in these parameters seen in the  \textsc{Nseries} galaxy mocks can be due to the combination of the following two effects. Compared to \textsc{Quijote}, \textsc{Nseries}  simulates biased tracers. Also, the true statistical errors in \textsc{Quijote} are the ones corresponding to the mean signal, which is that of a volume of 8000 (Gpc$h^{-1})^3$, while the statistical error of \textsc{Nseries} corresponds to a volume 100 times smaller. Hence, the bias we see in $P_{02}$ in \textsc{Nseries} could just be a $2\sigma$ statistical fluctuation.

In the right panel of Figure \ref{fig:Quijote_P02B02_fs8post}, we display the cosmological parameter fit for \textsc{Quijote} at redshifts $z=1,2$, with the $Z_2^\textrm{GEO}$ kernel fixed at the best-fit coefficients fitted at the corresponding redshifts (Table \ref{tab: fit_quij}). While the cosmology is broadly recovered, the accuracy is worse than at $z=0.5$  (Section \ref{sec: Quijresults}).  We attribute this effect in large part to the  power spectrum data-vector $P_{02}$, which is more affected by small-scales loss of power at $k\gtrsim0.12\,h\,{\rm Mpc}^{-1}$ than at lower redshift.

We further quantify the importance of developing and adopting a more  accurate model for the power spectrum in Table \ref{tab: deltachisq}. There we report the $\Delta \chi^2$ between the data vector for the best fit and that for the fiducial parameters for the three redshifts of interest in  \textsc{Quijote}, $z=0.5,1,2$. What we refer as ``fiducial'' are the parameters that are shown as a dashed line in Figures \ref{fig:Quijote_P02B02_fs8post} and \ref{fig:Quijote_P02B02_nog2poc_5parnorm}: namely, the $\sigma_8$ and $f$ values as computed by \textsc{Class} for the given cosmology and redshift; the bias parameters $b_1=1, b_2=0$; the shot-noise parameters $A_\textrm{P}=A_\textrm{B}=1$; 
the $\sigma_\textrm{P},\sigma_\textrm{B}$ are the (conditional\footnote{With the previous parameters fixed at the fiducial values.}) best fit to the simulations data. The table shows that the  shift in $\chi^2$ induced by $B_{02}$ is minuscule compared to that induced by $P_{02}$
considering the number of bispectrum configurations:  the power spectrum has a very strong
preference for the biased cosmological parameters and is driving the offsets seen in Figures \ref{fig:Quijote_P02B02_fs8post} and \ref{fig:Quijote_P02B02_nog2poc_5parnorm}. This is important, but goes beyond the scope of  this paper and  will be addressed elsewhere.

\begin{table}[htb]
\centering
\begin{tabular}{|l||c|c|c|}
\hline
\multicolumn{4}{|c|}{\textsc{Quijote} $\Delta \chi^2$ respect to fiducial}  \\\hline
Covariance terms     &  $z=0.5$              & $z=1$                   & $z=2$         \\
\hline
$P_{02}$        & -87.26 &  -286.07        & -295.90     \\ %
$B_{02}$     & -9.84  & -2.03         & -2.57        \\ %
$P_{02}+B_{02}$ & -174.98  &-478.86        &-481.52\\ 
Off-diagonal & -77.88  & -190.76        & -183.05\\ \hline
\end{tabular}

\caption{Difference in $\chi^2$ between the best-fit  and the fiducial cosmological parameters for the  \textsc{Quijote} simulations at $z=0.5,1,2$. The fiducial cosmological parameters are the $\sigma_8$ and $f$ corresponding to each redshift; $b_1=1, b_2=0, A_\textrm{P}=A_\textrm{B}=0$; and $\sigma_\textrm{P},\sigma_\textrm{B}$ obtained as the best fit for the full $P_{02}+B_{02}$ data-vector considering the remaining parameters fixed at their fiducial values. In each row we display the values corresponding to different parts of the data-vector, which correspond to different blocks of the covariance matrix: the $P_{02}$ block, the $B_{02}$ block, the full covariance matrix ($P_{02}+B_{02}$), and the off-diagonal contribution. The $B_{02}$ shift in $\chi^2$ is much smaller than that of the $P_{02}$ indicating that the power spectrum has a very strong preference for the biased cosmological parameters but not the bispectrum. This confirms that the power spectrum is driving the offsets seen in Figures \ref{fig:Quijote_P02B02_fs8post} and \ref{fig:Quijote_P02B02_nog2poc_5parnorm} (where for the lower redshift $z=0.5$ the offsets are much less pronounced). }
\label{tab: deltachisq}
\end{table}

\section{\textsc{Nseries} complementary results}
\label{App: Nseries}

We assess here the performance of the ``2-kernels'' extension of  Equation \ref{eq: Z2_geo} which modifies the $Z_2^{\rm GEO}$ kernels by introducing 5 extra free coefficients $\{g_1,g_2,g_3,g_4,g_5\}$:

\begin{align}
    Z_2^{\textrm{GEO}}\to Z_2^{SPT}\times\Big[(f_1+\ell g_1)&+(f_2+\ell g_2)\frac{\cos(\theta_\textrm{med})}{\cos(\theta_\textrm{max})}+(f_3+\ell g_3)\frac{\cos(\theta_\textrm{min})}{\cos(\theta_\textrm{max})}\nonumber\\
    &+(f_4+\ell g_4)\frac{A}{A_{\rm norm}}+(f_5+\ell g_5)\frac{A^2}{A_{\rm norm}}\Big],
    \label{eq: Z_2 2kern}
\end{align}
where $\ell=0$ for the bispectrum monopole and $\ell=1$ for the quadrupoles. 
The first set of parameters, $\{f_1,...,f_7,\sigma_B\}$, is adjusted to the $B_0$ \textsc{Quijote} measurements. Then, keeping their values fixed at the best fit, the $\{g_1,...,g_5\}$ parameters values  (which are only activated when $\ell\neq0$)  are fitted to  the $B_0+B_{200}$ \textsc{Quijote} measurements.

\begin{figure}[htb]
\centering 
\includegraphics[  width = \textwidth ]
{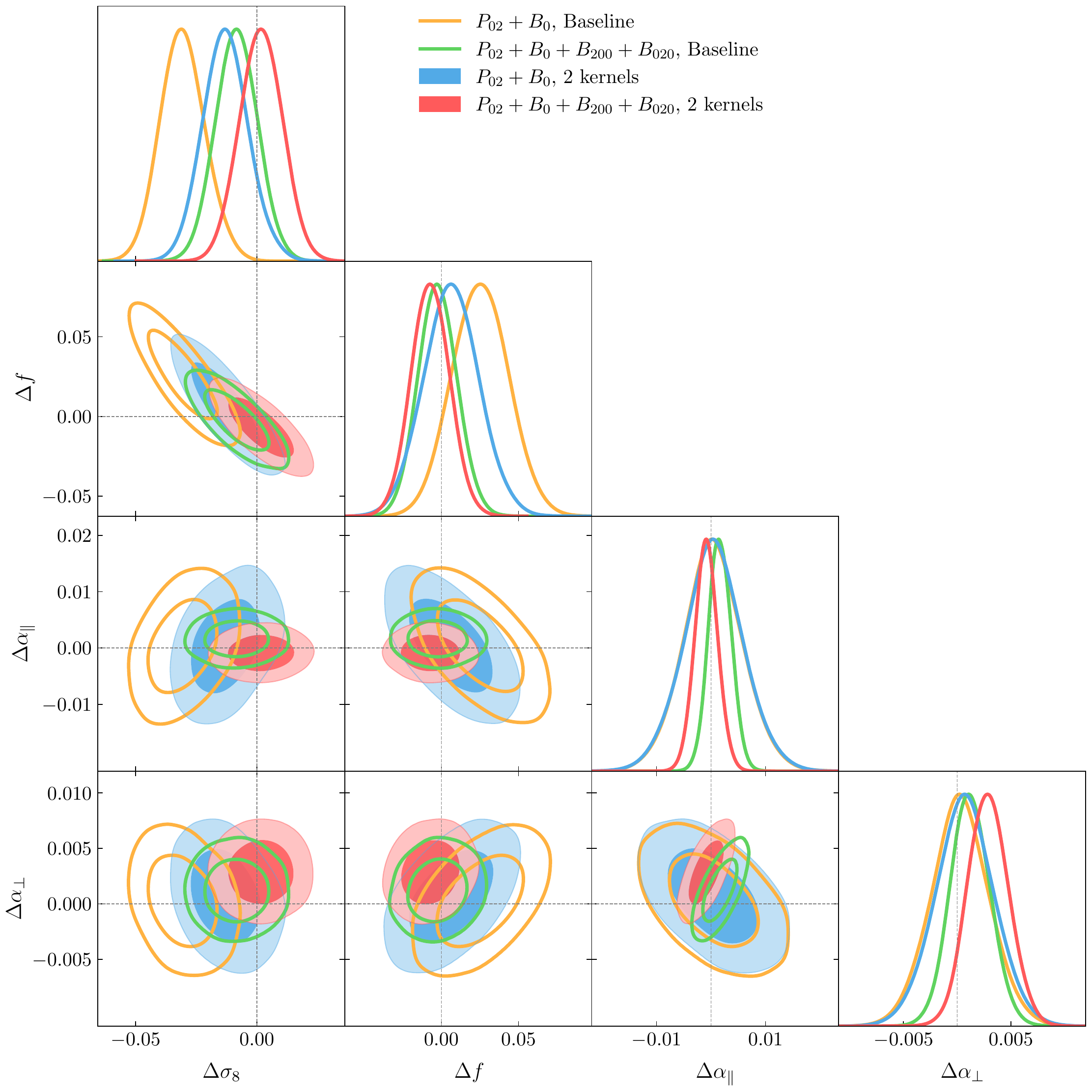}
\caption{Comparison of the approaches presented in Equation \ref{eq: Z2_geo} (baseline, being the same as in Figure \ref{fig:Nseries_P02B02_nog2poc_5parnorm}) and Equation \ref{eq: Z_2 2kern} (2 kernels) for the data-vectors $P_0+P_2+B_0$ and $P_0+P_2+B_0+B_{200}+B_{020}$.}
\label{fig:Nseries_P02B02_nog2poc_5parnorm12kern}
\end{figure}

In this way, the $f_i$ parameters are optimized for the bispectrum monopole, while the $g_i$ parameters quantify the discrepancy between the best functional form for the monopole and  that for the quadrupoles. As such, we should expect some  improvement in the accuracy of the constraints of cosmological parameters. 

As shown in Figure \ref{fig:Nseries_P02B02_nog2poc_5parnorm12kern}, the improvement is minimal for our data-vector of  interest, $P_{02}+B_{02}$. 
For this reason the 2-kernels extension of Equation \ref{eq: Z_2 2kern} is not considered in the main text of the paper.
This finding indicates that  our adopted  baseline model does not perform significantly better in some multipoles than in others.

Nonetheless, when using only the  $P_{02}+B_0$ data vector, the 2-kernels modification (i.e., fitting the eq.~\ref{eq: Z2_geo} kernel parameters only to  $B_0$) improves the recovery of the  individual $f$ and $\sigma_8$ parameters.%
We recognize that this modification can be suitable in cases where adding the quadrupoles is unfeasible and only the $P_{02}+B_0$ data vector is available. However, considering that, as shown in Section \ref{sec: fs8 disentangle}, the offset in the $f$ and $\sigma_8$ parameters is mainly caused by the power spectrum, the improvement seen in  the  2-kernels case for the  $P_{02}+B_0$ data-vector may be spurious.
Nevertheless, we provide the option to use the $B_0$-fitted $f_i$ parameters in the released code.

\begin{figure}[htb]
\centering 
\includegraphics[  width = 0.8\textwidth ]
{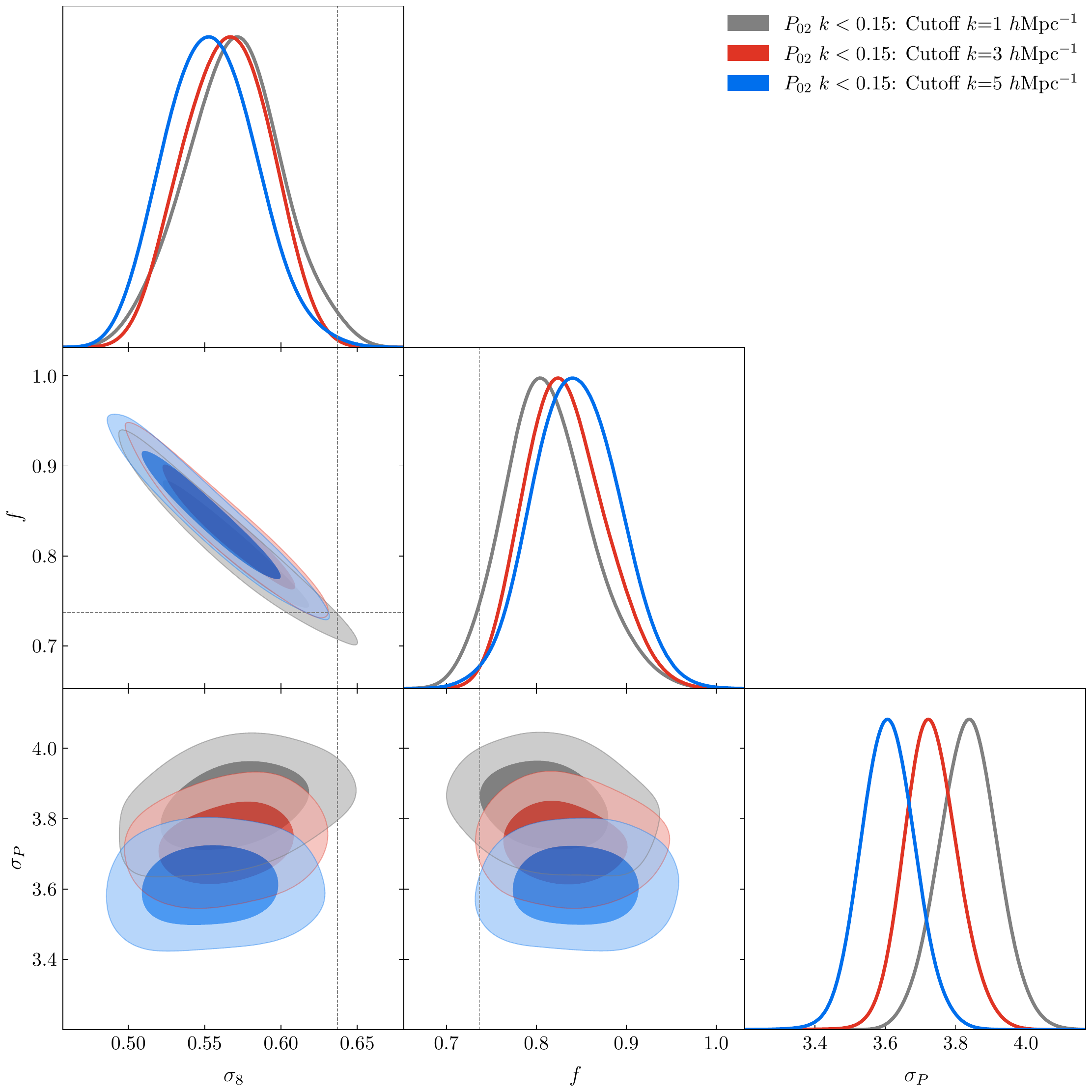}
\caption{Posterior distributions of the recovered cosmological parameters $\sigma_8,f$ and the nuisance parameter $\sigma_{\rm FoG}$, for the power spectrum data-vector in the \textsc{Nseries} simulations as a function of the cutoff $k$. This cutoff $k$ (not to be confused with the $k_\textrm{max}$ of the data-vector, which we set to 0.15 in this plot), is the integral upper limit for the loop correction terms of the power spectrum model. We observe how the offset featured in Figure \ref{fig:Nseries_P02B02_fs8post} is not ameliorated by a choice of a larger cutoff $k$; rather, this choice is absorbed in the nuisance parameter $\sigma_\textrm{P}$, leaving the rest of the parameters largely unaffected. }
\label{fig:Nseries_P02_kmaxint}
\end{figure}

Finally, we display in Figure \ref{fig:Nseries_P02_kmaxint} the behaviour of the $P_{02}$-recovered parameters $f,\sigma_8$ and the nuisance parameter $\sigma_P$, depending on the value of the cutoff $k$ of the integrals for the loop terms of the power spectrum. We clearly observe how the effect of increasing the cutoff $k$ is absorbed by the FoG nuisance parameter $\sigma_P$, thus discarding the  specific choice of cutoff $k$ as the source for the bias in disentangling $f$ and $\sigma_8$.

\bibliographystyle{ieeetr}
\bibliography{references}

\end{document}